\documentclass[proceedings]{JHEP37}
\usepackage{eurosym}
\usepackage{amssymb}
\usepackage{amsfonts}
\usepackage{amsmath,epsfig}
\usepackage{graphicx}

\newbox\mybox

\newcommand\fverb{\setbox\mybox=\hbox\bgroup\verb}
\newcommand\fverbdo{\egroup\medskip\noindent\fbox{\unhbox\mybox}\ }
\newcommand\fverbit{\egroup\item[\fbox{\unhbox\mybox}]}
\conference{Integrable nonlocal Hirota equations}
\abstract{We construct several new integrable systems corresponding to nonlocal versions of the Hirota equation, which is a particular example
of higher order nonlinear Schr\"{o}dinger equations. The integrability of the new models is established by providing their explicit forms of
Lax pairs or zero curvature conditions. The two compatibility equations arising in this construction are found to be related to each other either by a parity transformation $\mathcal{P}$, by a time 
reversal $\mathcal{T}$ or a $\mathcal{PT}$-transformation possibly combined with a conjugation. We construct explicit multi-soliton solutions for these models by employing Hirota's direct method 
as well as Darboux-Crum transformations. The nonlocal nature of these models allows for a modification of these solution procedures as the new systems also possess new types of solutions 
with different parameter dependence and different qualitative behaviour. The multi-soliton solutions are of varied type, being for instance nonlocal in space, nonlocal in time of time crystal type, 
regular with local structures either in time/space or of rogues wave type. }

\title{Integrable nonlocal Hirota equations}
\author{Julia Cen$^\bullet$, Francisco Correa$^\circ$ and Andreas Fring$%
^\bullet$ \\
$\bullet$ Department of Mathematics, City, University of London,\\
$\,\,$ Northampton Square, London EC1V 0HB, UK \\
$\circ$ Instituto de Ciencias F{\'{\i}}sicas y Matem{\'{a}}ticas,
Universidad Austral de Chile, \\
$\,\,$ Casilla 567, Valdivia, Chile\\
E-mail: julia.cen.1@city.ac.uk, francisco.correa@uach.cl, a.fring@city.ac.uk}

\input{tcilatex}
\begin{document}

\section{Introduction}

The nonlinear Schr\"{o}dinger equation (NLSE) \cite{shabat1972exact} is a
well studied prototypical nonlinear integrable system with many physical
applications, most notably in nonlinear optics where it describes the wave
propagation in Kerr type media, see e.g. \cite{agrawal2012fiber}, or plasma
physics \cite{shukla2010}. The main interest in the NLSE arises from the
fact that due its integrability it possesses solitonic wave solutions that
can be realized in form of optical pulses. While the NLSE provides a very
accurate description for the wave propagation of pulses in the picosecond
regime \cite{mollenauer1980}, experiments in the high-intensity and short
pulse subpicosecond, i.e. femtosecond, regime \cite{mitschke1986,gordon1986}
suggested for higher order corrections to be taken into account. Motivated
by these physical reasons, Kodama and Hasegawa \cite{kodama1987} proposed
the higher order nonlinear Schr\"{o}dinger equation (HNLSE) 
\begin{equation}
iq_{t}+\frac{1}{2}q_{xx}+\left\vert q\right\vert ^{2}q+i\varepsilon \left[
\alpha q_{xxx}+\beta \left\vert q\right\vert ^{2}q_{x}+\gamma q\left\vert
q\right\vert _{x}^{2}\right] =0,  \label{HNLSE}
\end{equation}%
with constants $\varepsilon ,\alpha ,\beta ,\gamma \in \mathbb{R}$. Besides
the NLSE for $\varepsilon =0$, four cases are known to be integrable. When
the ratio of the constants are taken to be $\alpha :\beta :\gamma =0:1:1$ or 
$\alpha :\beta :\gamma =0:1:0$ one obtains the derivative NLSE of type I 
\cite{anderson1983} and II \cite{chen1979}, respectively, which are in fact
related to each other by a dependent variable transformation \cite%
{wadati1983}. For $\alpha :\beta :\gamma =1:6:3$ one obtains the
Sasa-Satsuma equation \cite{sasa1991} and for $\alpha :\beta :\gamma =1:6:0$
the Hirota equation \cite{hirota1973exact}. Variations of the latter are the
subject of this manuscript.

We notice that the additional term in the HNLSE when compared to the NLSE,
i.e. (\ref{HNLSE}) for $\varepsilon =0$, shares the same $\mathcal{PT}$%
-symmetry with the NLSE, as it is invariant with respect to $\mathcal{PT}%
:x\rightarrow -x$, $t\rightarrow -t$, $i\rightarrow -i$, $q\rightarrow q$,
where $\mathcal{P}:x\rightarrow -x$ and $\mathcal{T}:$ $t\rightarrow -t$, $%
i\rightarrow -i$. Hence HNLSEs may also be viewed as $\mathcal{PT}$%
-symmetric extensions of the NLSE. Similarly as for many other $\mathcal{PT}$%
-symmetric nonlinear integrable systems \cite{fring2013pt}, various other $%
\mathcal{PT}$-symmetric generalizations have been proposed and investigated
by adding terms to the original equation, e.g. \cite%
{abdullaev2011solitons,alexeeva2012optical,konotop2016nonlinear}. A further
option, that will be important here, was explored by Ablowitz and Musslimani 
\cite{ablowitz2013,ablowitz2016} who identified a new class of nonlinear
integrable systems closely related to the NLSE by exploiting a hitherto
unexplored $\mathcal{PT}$-symmetry present in the zero curvature condition.
Exploring this option below for the Hirota equation will lead us to new
integrable systems with nonlocal properties.

Our manuscript is organized as follows: In section 2 we discuss the zero
curvature condition or AKNS-equation for the new class of integrable
systems. The solutions to these systems involve fields at different points
in space or time and reduce in certain limits to the standard Hirota
equation, so that we refer to them as \emph{nonlocal Hirota equations}. The
equations possess two types of solutions of qualitatively different
behaviour and parameter dependence. We identify the origin for this novel
feature within the context of Hirota's direct method as well as in the
application of Darboux-Crum transformations. At first we discuss these two
solution methods for the local Hirota equation in section 3. This will not
only serve as a benchmark for what follows, but we will also report new
solutions to these equations. In section 4-7 we construct and discuss the
solutions for the different types of new models. Our conclusions are stated
in section 8.

\section{Zero curvature equations for nonlocal Hirota equations}

The classical integrability of a model can be established by the Painlev\'{e}
test \cite{Pain1,Pain2} or the explicit construction of its Lax pair \cite%
{Lax} which is equivalent to the closely related zero curvature condition,
also referred to as AKNS-equation \cite{AKNS}. While the former is a mere
test, essentially just providing a yes or no answer to the question of
whether a model is integrable or not, the latter is more constructive and
constitutes a starting point for an explicit solution procedure. The
reformulations of the equation of motion of the model in terms of the zero
curvature condition allows for the construction of infinitely many conserved
charges, which is synonymous to the model being classically integrable. We
explore various symmetries in this reformulation that will lead us to new
types of models exhibiting novel features.

In general, the zero curvature condition for two operators $U$ and $V$ is
equivalent to two linear first order differential equations for an auxiliary
function $\Psi $%
\begin{equation}
\partial _{t}U-\partial _{x}V+\left[ U,V\right] =0\qquad \Leftrightarrow
\qquad \Psi _{t}=V\Psi \text{, }\Psi _{x}=U\Psi .  \label{ZC}
\end{equation}%
For a concrete model these equation have to hold up to the validity of the
equation of motion. When taking the matrix valued functions $U$ and $V$ to
be of the general form \ 
\begin{equation}
U=\left( 
\begin{array}{cc}
-i\lambda & q(x,t) \\ 
r(x,t) & i\lambda%
\end{array}%
\right) ,\qquad V=\left( 
\begin{array}{cc}
A(x,t) & B(x,t) \\ 
C(x,t) & -A(x,t)%
\end{array}%
\right) ,\qquad
\end{equation}%
involving the constant spectral parameter $\lambda $ and at this point
arbitrary functions $q(x,t)$ and $r(x,t)$, the zero curvature condition
holds when the matrix entries $A$, $B$ and $C$ satisfy the coupled equations%
\begin{eqnarray}
A_{x}(x,t) &=&q(x,t)C(x,t)-r(x,t)B(x,t),  \label{a} \\
B_{x}(x,t) &=&q_{t}(x,t)-2q(x,t)A(x,t)-2i\lambda B(x,t),  \label{b} \\
C_{x}(x,t) &=&r_{t}(x,t)+2r(x,t)A(x,t)+2i\lambda C(x,t).  \label{c}
\end{eqnarray}%
Suppressing now the explicit $x,t$-dependence of the functions involved, a
solution to the equations (\ref{a})-(\ref{c}) with arbitrary constants $%
\alpha $, $\beta $ is 
\begin{eqnarray}
A &=&-i\alpha qr-2i\alpha \lambda ^{2}+\beta \left( rq_{x}-qr_{x}-4i\lambda
^{3}-2i\lambda qr\right) , \\
B &=&i\alpha q_{x}+2\alpha \lambda q+\beta \left( 2q^{2}r-q_{xx}+2i\lambda
q_{x}+4\lambda ^{2}q\right) , \\
C &=&-i\alpha r_{x}+2\alpha \lambda r+\beta \left( 2qr^{2}-r_{xx}-2i\lambda
r_{x}+4\lambda ^{2}r\right) ,
\end{eqnarray}%
when $q(x,t)$ and $r(x,t)$ satisfy the two equations%
\begin{eqnarray}
q_{t}-i\alpha q_{xx}+2i\alpha q^{2}r+\beta \left[ q_{xxx}-6qrq_{x}\right]
&=&0,  \label{zero1} \\
r_{t}+i\alpha r_{xx}-2i\alpha qr^{2}+\beta \left( r_{xxx}-6qrr_{x}\right)
&=&0.  \label{zero2}
\end{eqnarray}%
Next one needs to make sure that these two equations are in fact compatible.
Adopting now from \cite{ablowitz2013,ablowitz2016} the general idea that has
been applied to the NLSE\ to the current setting we explore various choices
and alter the $x,t$-dependence in the functions $r$ and $q$. For convenience
we suppress the explicit functional dependence and absorb it instead into
the function's name by introducing the abbreviations%
\begin{equation}
q:=q(x,t)\text{,\quad }\tilde{q}:=q(-x,t)\text{,\quad }\hat{q}:=q(x,-t)\text{%
,\quad }\check{q}:=q(-x,-t).\text{\quad }
\end{equation}%
All six choices for $r(x,t)$ being equal to $\tilde{q}$, $\hat{q}$, $\check{q%
}$ or their complex conjugates $\tilde{q}^{\ast }$, $\hat{q}^{\ast }$, $%
\check{q}^{\ast }$ together with some specific adjustments for the constants 
$\alpha $ and $\beta $ are consistent for the two equations (\ref{zero1})
and (\ref{zero2}), thus giving rise to six new types of integrable models
that have not been explored so far. We will first list them and then study
their properties, in particular their solutions, in the next chapters.

\subparagraph{$\!\!\!\!\!\!\!\!\!\!\!\!\!\!\!\!$ The Hirota equation, a
conjugate pair, $r(x,t)=\protect\kappa q^{\ast }(x,t)$:\newline
}

$\!\!\!\!\!\!\!\!\!\!\!$ The standard choice to achieve compatibility
between (\ref{zero1}) and (\ref{zero2}) is to take $r(x,t)=\kappa q^{\ast
}(x,t)$ with $\kappa =1$. Here we allow $\kappa \in \mathbb{R}$, such that
the equations acquire the forms%
\begin{eqnarray}
iq_{t} &=&\!-\alpha \left( q_{xx}-2\kappa \left\vert q\right\vert
^{2}q\right) \!-i\!\beta \!\left( q_{xxx}-6\kappa \left\vert q\right\vert
^{2}q_{x}\right) ,  \label{1} \\
-iq_{t}^{\ast } &=&-\alpha \left( q_{xx}^{\ast }-2\kappa \left\vert
q\right\vert ^{2}q^{\ast }\right) \!+i\!\beta \!\left( q_{xxx}^{\ast
}-6\kappa \left\vert q\right\vert ^{2}q_{x}^{\ast }\right) .\quad
\;\;\;\;\;\,  \label{2}
\end{eqnarray}%
Equation (\ref{1}) is the known Hirota equation \cite{hirota1973exact}.
Taking in (\ref{1}) $\kappa =1$, $\alpha \rightarrow 1/2$ and $\beta
\rightarrow \varepsilon $ we obtain the HNLSE (\ref{HNLSE}) when setting $%
\alpha \rightarrow 1$, $\beta \rightarrow 6$, $\gamma \rightarrow 0$ in
there. For $\alpha ,\beta ,\kappa \in \mathbb{R}$ equation (\ref{2}) is its
complex conjugate, respectively, i.e. (\ref{2})$^{\ast }=$(\ref{1}). When $%
\beta \rightarrow 0$ equation (\ref{1}) reduces to the NLSE with conjugate (%
\ref{2}) and for $\alpha \rightarrow 0$ equation (\ref{1}) reduces to the
complex modified Korteweg de-Vries with conjugate (\ref{2}). The
aforementioned $\mathcal{PT}$-symmetry is preserved in these equations.

\subparagraph{$\!\!\!\!\!\!\!\!\!\!\!\!\!\!\!\!$ A parity transformed
conjugate pair, $r(x,t)=\protect\kappa q^{\ast }(-x,t)$:\newline
}

$\!\!\!\!\!\!\!\!\!\!\!$ Taking now $r(x,t)=\kappa \tilde{q}^{\ast }$ with $%
\kappa \in \mathbb{R}$ together with $\beta =i\delta $, $\alpha ,\delta \in 
\mathbb{R}$, the equations (\ref{zero1}) and (\ref{zero2}) become 
\begin{eqnarray}
\!iq_{t}\! &=&\!-\alpha \left[ q_{xx}-2\kappa \tilde{q}^{\ast }q^{2}\right]
\!+\delta \!\left[ q_{xxx}-6\kappa q\tilde{q}^{\ast }q_{x}\right] ,
\label{new1} \\
-\!i\tilde{q}_{t}^{\ast } &=&-\!\alpha \left[ \tilde{q}_{xx}^{\ast }-2\kappa
q(\tilde{q}^{\ast })^{2}\right] \!-\delta \!\left( \tilde{q}_{xxx}^{\ast
}-6\kappa \tilde{q}^{\ast }q\tilde{q}_{x}^{\ast }\right) .\quad ~~~
\label{new2}
\end{eqnarray}%
We observe that equation (\ref{new1}) is the parity transformed conjugate of
equation (\ref{new2}), i.e. $\mathcal{P}$(\ref{new1})$^{\ast }=$(\ref{new2}%
). We also notice that a consequence of the introduction of the nonlocality
is that the aforementioned $\mathcal{PT}$-symmetry has been broken.

\subparagraph{$\!\!\!\!\!\!\!\!\!\!\!\!\!\!\!\!$ A time-reversed pair, $%
r(x,t)=\protect\kappa q^{\ast }(x,-t)$:\newline
}

$\!\!\!\!\!\!\!\!\!\!\!$ Choosing $r(x,t)=\kappa \hat{q}^{\ast }$ with $%
\kappa \in \mathbb{R}$ and $\alpha =i\hat{\delta}$, $\beta =i\delta $, $\hat{%
\delta},\delta \in \mathbb{R}$ we obtain from equations (\ref{zero1}) and (%
\ref{zero2}) the pair 
\begin{eqnarray}
\!iq_{t}\! &=&\!-i\hat{\delta}\left[ q_{xx}-2\kappa \hat{q}^{\ast }q^{2}%
\right] \!+\delta \!\left[ q_{xxx}-6\kappa q\hat{q}^{\ast }q_{x}\right] ,
\label{m1} \\
\!i\hat{q}_{t}^{\ast } &=&\!i\hat{\delta}\left[ \hat{q}_{xx}^{\ast }-2\kappa
q(\hat{q}^{\ast })^{2}\right] \!+\delta \!\left( \hat{q}_{xxx}^{\ast
}-6\kappa \hat{q}^{\ast }q\hat{q}_{x}^{\ast }\right) .\quad ~~~  \label{m2}
\end{eqnarray}%
Recalling here that the time-reversal map includes a conjugation, such that $%
\mathcal{T}:q\rightarrow \hat{q}^{\ast },i\rightarrow -i$, we observe that (%
\ref{m1}) is the time-reversed of equations (\ref{m2}), i.e. $\mathcal{T}$(%
\ref{m2})$=$(\ref{m1}). The $\mathcal{PT}$-symmetry is also broken in this
case.

\subparagraph{$\!\!\!\!\!\!\!\!\!\!\!\!\!\!\!\!$ A $\mathcal{PT}$-symmetric
pair, $r(x,t)=\protect\kappa q^{\ast }(-x,-t)$: \newline
}

$\!\!\!\!\!\!\!\!\!\!\!$ For the choice $r(x,t)=\kappa \check{q}^{\ast }$
with $\kappa \in \mathbb{R}$ and $\alpha =i\check{\delta}$, $\check{\delta}%
,\beta \in \mathbb{R}$ the equations (\ref{zero1}) and (\ref{zero2}) become 
\begin{eqnarray}
\!q_{t}\! &=&\!-\check{\delta}\left[ q_{xx}-2\kappa \check{q}^{\ast }q^{2}%
\right] \!-\beta \!\left[ q_{xxx}-6\kappa q\check{q}^{\ast }q_{x}\right] ,
\label{pt1} \\
\!-\check{q}_{t}^{\ast } &=&-\!\check{\delta}\left[ \check{q}_{xx}^{\ast
}-2\kappa q(\check{q}^{\ast })^{2}\right] \!+\beta \!\left( \check{q}%
_{xxx}^{\ast }-6\kappa \check{q}^{\ast }q\check{q}_{x}^{\ast }\right) .\quad
~~~  \label{pt2}
\end{eqnarray}%
We observe that the overall constant $i$ has cancelled out and the two
equations are transformed into each other by means of a $\mathcal{PT}$%
-symmetry transformation $\mathcal{PT}$(\ref{pt2})$=$(\ref{pt1}). Thus,
while the $\mathcal{PT}$-symmetry for the equations (\ref{pt1}) is broken,
the two equations are transformed into each other by that symmetry.

\subparagraph{$\!\!\!\!\!\!\!\!\!\!\!\!\!\!\!\!$ A real parity transformed
conjugate pair, $r(x,t)=\protect\kappa q(-x,t)$:\newline
}

$\!\!\!\!\!\!\!\!\!\!\!$ We may also choose $q(x,t)$ to be real. For $%
r(x,t)=\kappa \tilde{q}$ with $\kappa ,\tilde{q}\in \mathbb{R}$ and $\beta
=i\delta $, $\alpha ,\delta \in \mathbb{R}$, the equations (\ref{zero1}) and
(\ref{zero2}) acquire the forms 
\begin{eqnarray}
\!iq_{t}\! &=&\!-\alpha \left[ q_{xx}-2\kappa \tilde{q}q^{2}\right]
\!+\delta \!\left[ q_{xxx}-6\kappa q\tilde{q}q_{x}\right] ,  \label{pr1} \\
\!-i\tilde{q}_{t} &=&\!-\alpha \left[ \tilde{q}_{xx}-2\kappa q\tilde{q}^{2}%
\right] \!-\delta \!\left( \tilde{q}_{xxx}-6\kappa \tilde{q}q\tilde{q}%
_{x}\right) .\quad ~~~  \label{pr2}
\end{eqnarray}%
Just as their complex variants (\ref{zero1}) and (\ref{zero2}), also the
equations (\ref{pr2}) and (\ref{pr1}) are related to each other by
conjugation and a parity transformation (\ref{new2}), i.e. $\mathcal{P}$(\ref%
{pr2})$^{\ast }=$(\ref{pr1}). However, the restriction to real values for $%
q(x,t)$ makes these equations less interesting as $q$ becomes static, which
simply follows from the fact that the left hand sides of (\ref{pr1}) and (%
\ref{pr2}) are complex valued, whereas the right hand sides are real valued.

\subparagraph{$\!\!\!\!\!\!\!\!\!\!\!\!\!\!\!\!$ A real time-reversed pair, $%
r(x,t)=\protect\kappa q(x,-t)$:\newline
}

$\!\!\!\!\!\!\!\!\!\!\!$ For $r(x,t)=\kappa \hat{q}$ with $\kappa ,\hat{q}%
\in \mathbb{R}$ and $\alpha =i\hat{\delta}$, $\beta =i\delta $, $\hat{\delta}%
,\delta \in \mathbb{R}$ we obtain from (\ref{zero1}) and (\ref{zero2}) 
\begin{eqnarray}
\!iq_{t}\! &=&\!-i\hat{\delta}\left[ q_{xx}-2\kappa \hat{q}^{\ast }q^{2}%
\right] \!+\delta \!\left[ q_{xxx}-6\kappa q\hat{q}^{\ast }q_{x}\right] ,
\label{T1} \\
\!\!i\hat{q}_{t}^{\ast } &=&\!i\hat{\delta}\left[ \hat{q}_{xx}^{\ast
}-2\kappa q(\hat{q}^{\ast })^{2}\right] \!+\delta \!\left( \hat{q}%
_{xxx}^{\ast }-6\kappa \hat{q}^{\ast }q\hat{q}_{x}^{\ast }\right) .\quad ~~~
\label{T2}
\end{eqnarray}%
Again we observe the same behaviour as in the complex variant, namely that
the two equations (\ref{T1}) and (\ref{T2}) become their time-reversed
counterparts, i.e. $\mathcal{T}$(\ref{T2})$=$(\ref{T1}). However, as a
consequence of $q$ being real these equations simply become the time-reverse
NLSE with the additional constraint $q_{xxx}=\pm 6q\hat{q}q_{x}$.

\subparagraph{$\!\!\!\!\!\!\!\!\!\!\!\!\!\!\!\!$ A conjugate $\mathcal{PT}$%
-symmetric pair, $r(x,t)=\protect\kappa q(-x,-t)$:\newline
}

$\!\!\!\!\!\!\!\!\!\!\!$ For our final choice $r(x,t)=\kappa \check{q}$ we
have no restriction on the constants, i.e. $\kappa ,\alpha ,\beta \in 
\mathbb{C}$, the equations (\ref{zero1}) and (\ref{zero2}) become 
\begin{eqnarray}
\!q_{t}\! &=&\!i\alpha \left[ q_{xx}-2\kappa \check{q}q^{2}\right] \!-\beta
\!\left[ q_{xxx}-6\kappa q\check{q}q_{x}\right] ,  \label{F1} \\
\!-\check{q}_{t} &=&\!i\alpha \left[ \check{q}_{xx}-2\kappa q\check{q}^{2}%
\right] \!+\beta \!\left( \check{q}_{xxx}-6\kappa \check{q}q\check{q}%
_{x}\right) .\quad ~~~  \label{F2}
\end{eqnarray}%
These two equations are transformed into each other by means of a $\mathcal{%
PT}$-symmetry transformation and a conjugation $\mathcal{PT}$(\ref{F2})$%
^{\ast }=$(\ref{F1}). A comment is in order here to avoid confusion. Since a
conjugation is included into the $\mathcal{T}$-operator, the additional
conjugation of (\ref{F1}) when transformed into (\ref{F2}) means that we
simply carry out $x\rightarrow -x$ and $t\rightarrow -t$. \ 

The paired up equations (\ref{1})-(\ref{F2}) are all new integrable systems.
Let us now discuss solutions and properties of these equations. Since the
two equations in each pair are related to each other by a well identified
symmetry transformation involving combinations of conjugation, reflections
in space and reversal in time, it suffices to focus on just one of the
equations.

\section{The local Hirota equations, a conjugate pair}

Even though the standard Hirota equation \cite{hirota1973exact} and many of
its solutions are known, we briefly recall the solution procedure and some
of its properties. This will serve as a benchmark that allows us to point
out the novelties of the nonlocal equations. We will also report some new
solutions. As mentioned, in this case the two equations (\ref{zero1}) and (%
\ref{zero2}) are compatible with the choice $r(x,t)=\kappa q^{\ast }(x,t)$.

\subsection{Hirota's direct method}

We start by presenting the bilinearisation for the equations (\ref{1}) and (%
\ref{2}), focusing on (\ref{1}) for the above mentioned reasons. Factorizing
the Hirota field as $q(x,t)=g(x,t)/f(x,t)$, with the assumptions $g(x,t)\in 
\mathbb{C}$, $f(x,t)\in \mathbb{R}$, we find the identify%
\begin{eqnarray}
&&f^{3}\left[ iq_{t}+\alpha q_{xx}-2\kappa \alpha \left\vert q\right\vert
^{2}q+i\!\beta \!\left( q_{xxx}-6\kappa \left\vert q\right\vert
^{2}q_{x}\right) \right] =  \label{BINLSE} \\
&&f\left[ iD_{t}g\cdot f+\alpha D_{x}^{2}g\cdot f+i\beta D_{x}^{3}g\cdot f%
\right] +\left[ 3i\beta \left( \frac{g}{f}f_{x}-g_{x}\right) -\alpha g\right]
\left[ D_{x}^{2}f\cdot f+2\kappa \left\vert g\right\vert ^{2}\right] .~~~ 
\notag
\end{eqnarray}%
The operators $D_{x}$, $D_{t}$ denote the standard Hirota derivatives \cite%
{hirota2004direct} defined by a Leibniz rule with alternating signs%
\begin{equation}
D_{x}^{n}f\cdot g=\sum\limits_{k=0}^{n}\left( 
\begin{array}{l}
n \\ 
k%
\end{array}%
\right) (-1)^{k}\frac{\partial ^{n-k}}{\partial x^{n-k}}f(x)\frac{\partial
^{k}}{\partial x^{k}}g(x).
\end{equation}%
Here we use the explicit expressions for $D_{t}f\cdot g=f_{t}g-fg_{t}$, $%
D_{x}^{2}f\cdot g=f_{xx}g-2f_{x}g_{x}+fg_{xx}$ and $D_{x}^{3}f\cdot
g=f_{xxx}g-3f_{xx}g_{x}+3f_{x}g_{xx}-fg_{xxx}$. The equation (\ref{BINLSE})
is still trilinear in the functions $f$, $g~$and not yet bilinear as
required for the applicability of Hirota's direct method. However, the left
hand side vanishes when the Hirota equation (\ref{1}) holds and the right
hand side becomes zero when the two bilinear equations%
\begin{eqnarray}
iD_{t}g\cdot f+\alpha D_{x}^{2}g\cdot f+i\beta D_{x}^{3}g\cdot f &=&0,
\label{H1} \\
D_{x}^{2}f\cdot f &=&-2\kappa \left\vert g\right\vert ^{2},  \label{H2}
\end{eqnarray}%
are satisfied. For $\alpha =1/2$ and $\kappa =-1$ they correspond to the
equations reported in \cite{liu2008soliton}. When $\beta \rightarrow 0$ the
equations (\ref{H1}) and (\ref{H2}) reduce to the bilinear form
corresponding to the NLSE \cite{hirota1973exact,hietarinta4}. The well known
virtue of this formulation is that the bilinear forms can be solved
systematically by using the formal power series expansions%
\begin{equation}
f(x,t)=\dsum\nolimits_{k=0}^{\infty }\varepsilon ^{2k}f_{2k}(x,t),\quad 
\text{and\quad }g(x,t)=\dsum\nolimits_{k=1}^{\infty }\varepsilon
^{2k-1}g_{2k-1}(x,t).  \label{expfg}
\end{equation}%
Solving recursively the equations that result when setting the coefficients
of each order in $\varepsilon $ to zero, one obtains different types of
solutions corresponding to $n$-soliton solutions with $n$ depending on the
order of expansion. A further well known virtue of Hirota's direct method is
the remarkable fact that the quantity $\varepsilon $ is only a formal
parameter and can be set to any value. Moreover, despite the fact that
initially the Ansatz for the solutions appear to be perturbative, the
truncated expansions become exact when setting $f_{k}(x,t)=$ $g_{n}(x,t)=0$
for $k>\ell $, $n>m$, for certain values of $\ell $ and $m$. We will see
below that in the nonlocal case we have the new option to weaken this
condition which then leads to additional new types of solutions.

In the manner just described the general one-soliton solution can be found
by using the truncated expansions $f(x,t)=1+\varepsilon ^{2}f_{2}(x,t)$ and $%
g(x,t)=\varepsilon g_{1}(x,t)$ in (\ref{H1}) and (\ref{H2}). Setting $%
\varepsilon =1$ without loss of generality, we obtain the local solution 
\begin{equation}
q_{\text{l}}^{(1)}(x,t)=\frac{g_{1}(x,t)}{1+f_{2}(x,t)},~~~~\ \ \ \ \ \ 
\text{with \ }g_{1}=\lambda \tau _{\mu ,\gamma }\text{,~~~}f_{2}(x,t)=\frac{%
-\kappa \left\vert \lambda \right\vert ^{2}}{(\mu +\mu ^{\ast })^{2}}%
\left\vert \tau _{\mu ,\gamma }\right\vert ^{2},  \label{oneS}
\end{equation}%
with constants $\mu $,$\gamma $,$\lambda \in \mathbb{C}$ and function%
\begin{equation}
\tau _{\mu ,\gamma }(x,t):=e^{\mu x+\mu ^{2}(i\alpha -\beta \mu )t+\gamma }.
\end{equation}%
We observe that for real parameter $\alpha $ the presence of the deformation
parameter $\beta $ changes drastically the overall qualitative behaviour of
the wave. When it is vanishing, that is in the case of the NLSE, the
solution is simply a standing wave that changes its amplitude as a function
of time. However, when $\beta $ is switched on the solutions of the full
Hirota equation displays a qualitatively different behaviour than the one
for the NLSE as the wave starts to move with a speed $v=\beta \mu ^{2}$.

\FIGURE{ \epsfig{file=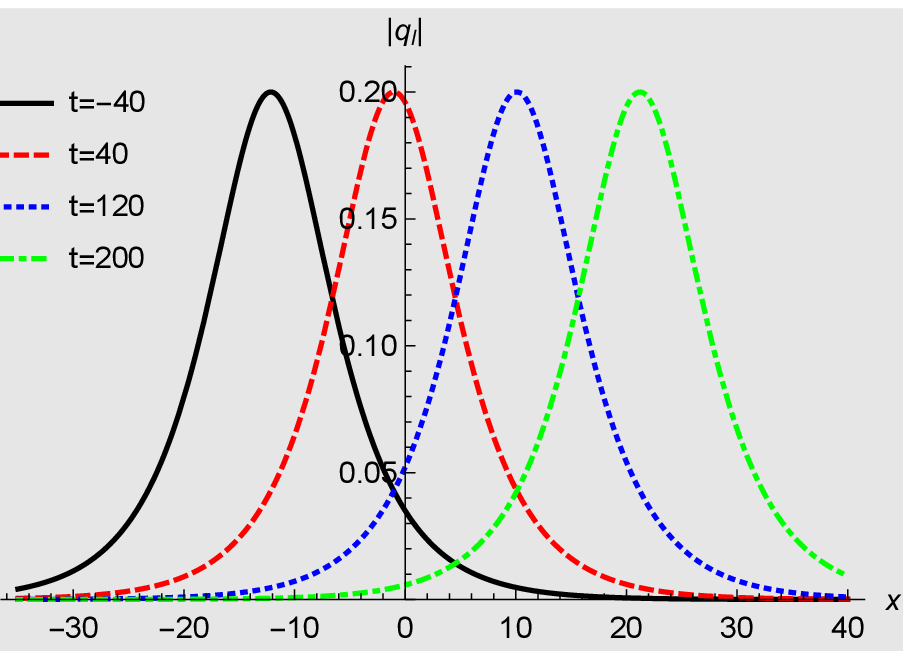,width=7.25cm} \epsfig{file=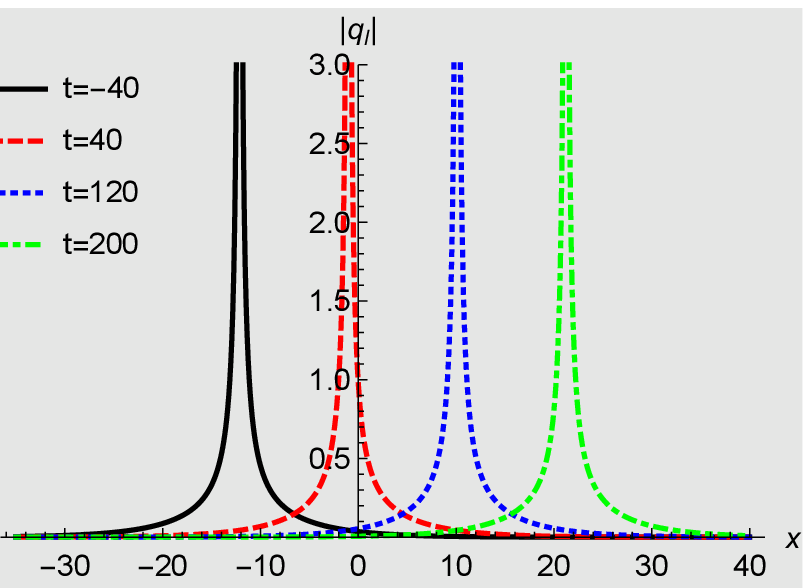,width=7.25cm}
\caption{Regular and singular  one-soliton solutions (\ref{oneS})  for the local Hirota equations (\ref{1}) at different times for $\alpha = 0.5$, $\beta=0.7$, 
        $\gamma=0.4+i0.3$, $\mu=0.2+i0.3$, $\lambda=1$ for $\kappa=-1$ and $\kappa=1$ in the left and right panel, respectively.}
        \label{Fig1}}

From figure \ref{Fig1} we also observe that for $\kappa =1$ the solution (%
\ref{oneS}) develops a singularity. Even though these cusp solutions have
possible applications \cite{eggers2001air} and are interesting in their own
right, we will often just focus on the equation for $\kappa =-1$ in what
follows, since apart from the overall sign the actual value of $\kappa $ is
irrelevant as it can be absorbed into $\lambda $.

To obtain the two-soliton solution we need to go two orders further in the
expansion (\ref{expfg}) and use $f(x,t)=1+\varepsilon
^{2}f_{2}(x,t)+\varepsilon ^{4}f_{4}(x,t)$, $g(x,t)=\varepsilon
g_{1}(x,t)+\varepsilon ^{3}g_{3}(x,t)$ in the bilinear equations (\ref{H1}),
(\ref{H2}). Setting $\varepsilon =1$ we obtain the two-soliton solution%
\begin{equation}
q_{\text{l}}^{(2)}(x,t)=\frac{g_{1}(x,t)+g_{3}(x,t)}{1+f_{2}(x,t)+f_{4}(x,t)}
\label{twoS}
\end{equation}%
with%
\begin{eqnarray}
g_{1} &=&\tau _{\mu ,\gamma }+\tau _{\nu ,\delta }, \\
g_{3} &=&\frac{\left( \mu -\nu \right) ^{2}}{\left( \mu +\mu ^{\ast }\right)
^{2}\left( \nu +\mu ^{\ast }\right) ^{2}}\tau _{\nu ,\delta }\left\vert \tau
_{\mu ,\gamma }\right\vert ^{2}+\frac{\left( \mu -\nu \right) ^{2}}{\left(
\mu +\nu ^{\ast }\right) ^{2}\left( \nu +\nu ^{\ast }\right) ^{2}}\tau _{\mu
,\gamma }\left\vert \tau _{\nu ,\delta }\right\vert ^{2}, \\
f_{2} &=&\frac{\left\vert \tau _{\mu ,\gamma }\right\vert ^{2}}{\left( \mu
+\mu ^{\ast }\right) ^{2}}+\frac{\tau _{\nu ,\delta }\tau _{\mu ,\gamma
}^{\ast }}{\left( \nu +\mu ^{\ast }\right) ^{2}}+\frac{\tau _{\mu ,\gamma
}\tau _{\nu ,\delta }^{\ast }}{\left( \mu +\nu ^{\ast }\right) ^{2}}+\frac{%
\left\vert \tau _{\nu ,\delta }\right\vert ^{2}}{\left( \nu +\nu ^{\ast
}\right) ^{2}}, \\
f_{4} &=&\frac{\left( \mu -\nu \right) ^{2}\left( \mu ^{\ast }-\nu ^{\ast
}\right) ^{2}}{\left( \mu +\mu ^{\ast }\right) ^{2}\left( \nu +\mu ^{\ast
}\right) ^{2}\left( \mu +\nu ^{\ast }\right) ^{2}\left( \nu +\nu ^{\ast
}\right) ^{2}}\left\vert \tau _{\mu ,\gamma }\right\vert ^{2}\left\vert \tau
_{\nu ,\delta }\right\vert ^{2}.
\end{eqnarray}

In comparison with the NLSE the Hirota equation exhibits a more varied
behaviour due to the presence of the additional parameter $\beta $. In
figure \ref{Fig2} we display a two-soliton composed of a fast one-soliton
overtaking a slower one. For a complex choice of the spectral and shift
parameter this behaviour is changed into a head-on collision of two
one-solitons.

\FIGURE{ \epsfig{file=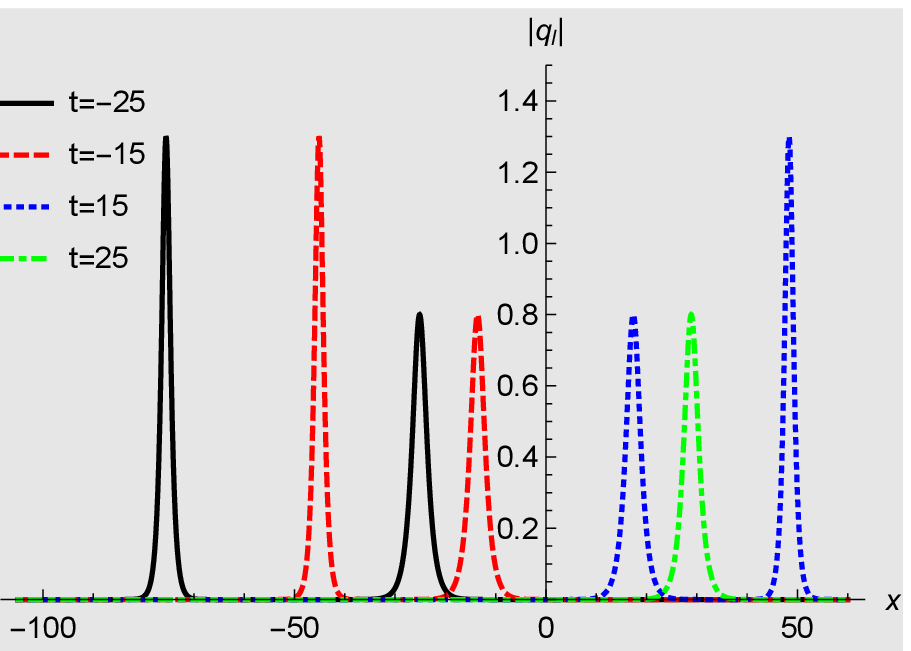,width=7.25cm} \epsfig{file=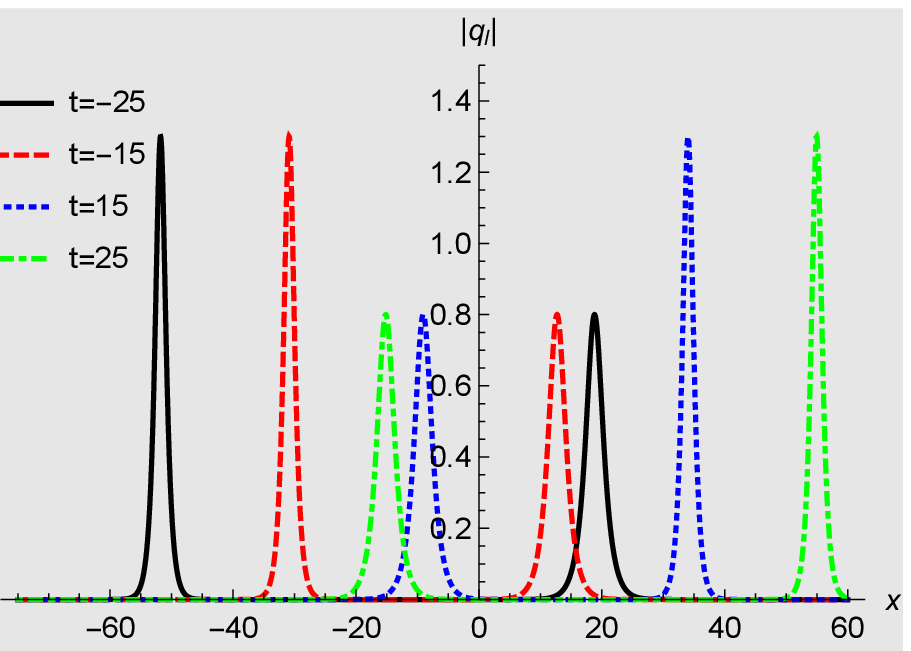,width=7.25cm}
\caption{Modulus of the two-soliton solutions (\ref{twoS})  for the local Hirota equations (\ref{1})  at different times for $\alpha = 0.4$, $\beta=1.8$, 
        $\gamma=0.3$, $\delta=0.4$, $\mu=1.3$, $\nu=0.8$,  $\lambda=1$,  $\kappa=-1$ displaying a faster soliton overtaking a slower one  (left panel) and  $\gamma=0.3+i0.1$, $\delta=0.4+i0.7$, $\mu=1.3+i0.5$, $\nu=0.8+i0.65$,  $\lambda=1$, $\kappa=1$ displaying a head-on collision (right panel).}
        \label{Fig2}}

More striking is the previously not pointed out possibility that within the
two-soliton solution one of the one-solitons contributions can be made
static by a suitable parameter choice. We observe in figure \ref{Fig3} that
the static soliton can be seen as a defect. The only effect of the
scattering is the usual slight displacement or time-delay depending on the
reference frame.

\FIGURE{ \epsfig{file=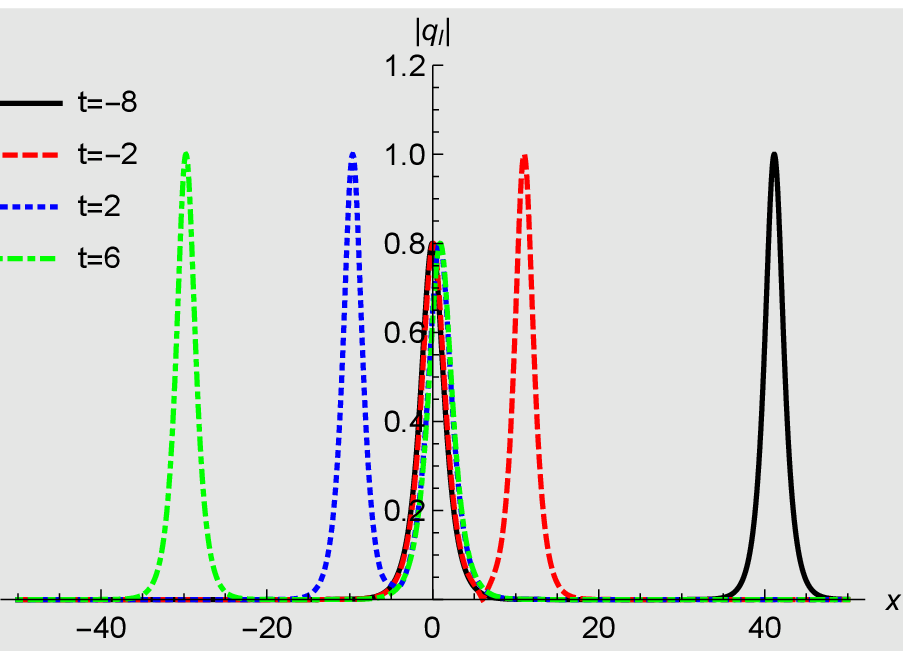,width=7.25cm}  \epsfig{file=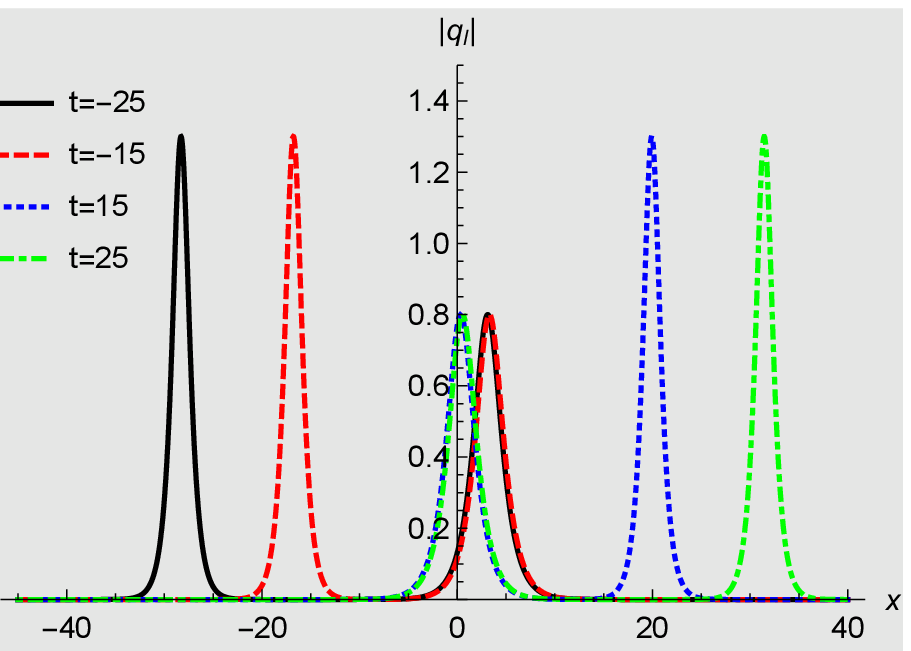,width=7.25cm}
\caption{Modulus of the two-soliton solutions (\ref{twoS}) for the local Hirota equations (\ref{1}) at different times for $\alpha = 0.4$, $\beta=0.8$, 
        $\gamma=0.3+i0.1$, $\delta=0.4+i0.7$, $\lambda=1$,  $\kappa=-1$ with a left moving one-soliton with $\mu=1.-i 1.4$, $\nu=0.8+i0.65$  (left panel) and a right moving one-soliton with
         $\mu=1.3+i0.5$, $\nu=0.8+i0.65$, (right panel) scattering with a static one-solition acting as a defect.}
        \label{Fig3}}

\subsection{Darboux-Crum transformations}

Alternatively, the solutions of the Hirota equation can also be constructed
following the Darboux-Crum transformation scheme \cite%
{darboux,crum,matveevdarboux}. At first we will keep our discussion very
general by leaving the functions $q$ and $r$ generic without specifying the
different choices for $r$ and consider those concrete scenarios in the next
sections.

Generally speaking, Darboux transformations relate two different Hamiltonian
systems by means of an intertwining relation \cite%
{darboux,crum,matveevdarboux}. The iteration of this procedure to a sequence
of Hamiltonian systems is usually referred to as the Darboux-Crum
transformation scheme. In the present case we can convert one of the AKNS
equations into an eigenvalue equation and thus identify a Hamiltonian of
Dirac type. Taking $\Psi $ to be a two-dimensional vector we obtain 
\begin{equation}
\Psi =\left( 
\begin{array}{c}
\varphi \\ 
\phi%
\end{array}%
\right) ,~~~~~~\quad \Psi _{x}=U\Psi ~\implies \quad 
\begin{array}{c}
-i\varphi _{x}+iq\phi =-\lambda \varphi \\ 
i\phi _{x}-ir\varphi =-\lambda \phi%
\end{array}%
\,.  \label{psi1}
\end{equation}%
Comparing with the eigenvalue equation $H\Psi (\lambda )=-\lambda \Psi
(\lambda )$, we read off the Hamiltonian 
\begin{equation}
H=\left( 
\begin{array}{cc}
-i\partial _{x} & iq \\ 
-ir & i\partial _{x}%
\end{array}%
\right) \,=-i\sigma _{3}\partial _{x}+V,
\end{equation}%
from (\ref{psi1}), with $\sigma _{3}$ denoting a standard Pauli matrix. Next
we seek to relate this Hamiltonian, together with its eigenfunctions, to a
set of new Hamiltonians of similar structure 
\begin{equation}
H_{n}=\left( 
\begin{array}{cc}
-i\partial _{x} & iq_{n} \\ 
-ir_{n} & i\partial _{x}%
\end{array}%
\right) \,=-i\sigma _{3}\partial _{x}+V_{n},  \label{Hn}
\end{equation}%
satisfying $H_{n}\Psi _{n}(\lambda )=-\lambda \Psi _{n}(\lambda )$ for $n\in 
\mathbb{N}$. By construction the new Hamiltonians are designed in such a way
that the $q_{n}$ and $r_{n}$ satisfy the two equations resulting from the
zero curvature condition with spectral parameters $\lambda _{n}$ and are
therefore also solutions to our nonlinear wave equations. Let us next
discuss how to obtain them by employing the Darboux-Crum transformation
scheme for Dirac Hamiltonians as discussed in \cite{nieto2003,correa2017}.
The key assumption is that the different Hamiltonians are recursively
related to each other by intertwining relations 
\begin{equation}
L_{n}H_{n-1}=H_{n}L_{n},  \label{inter}
\end{equation}%
with intertwining operators $L_{n}$. Identifying $H_{0}=H$, the iteration of
the equations (\ref{inter}) lead to the relation $\mathcal{L}_{n}H=H_{n}%
\mathcal{L}_{n}$ with $\mathcal{L}_{n}:=L_{n}L_{n-1}\ldots L_{1}$. It is
also easily verified that the wavefunctions at each level of iteration are
simply $\Psi _{n}(\lambda )=\mathcal{L}_{n}\Psi (\lambda )$.

Next we discuss how to obtain the intertwining operators and the potentials.
We start with equation (\ref{inter}) for $n=1$ and assume the intertwining
operator to be of the general form $L_{1}:=$ $\mathbb{I}\partial _{x}+B$.
Upon substituting $H,H_{1}$ and $L_{1}$the intertwining relation yields the
two equations%
\begin{equation}
V_{1}=V-i\left[ B,\sigma _{3}\right] ,\qquad \text{and\qquad }%
V_{x}+BV-V_{1}B+i\sigma _{3}B_{x}=0.  \label{22}
\end{equation}%
Taking next $B=-U_{x}U^{-1}$, as suggested in \cite{nieto2003}, and
substituting the first equation in (\ref{22}) into the second, the latter
becomes equivalent to%
\begin{equation}
\left( U^{-1}HU\right) _{x}=0.
\end{equation}%
Integrating this equation leads to $HU=U\Lambda $ with $\Lambda $ containing
the integration constants. This equation has now become formally equivalent
to the Schr\"{o}dinger equation with the difference that $U$ is a matrix.
Taking $\limfunc{diag}\Lambda =(-\lambda _{1},-\lambda _{2})$ this equation
is solved by $U=\left( \Psi (\lambda _{1}),\Psi (\lambda _{2})\right)
=:U_{1} $ and thus we have found%
\begin{equation}
L_{1}=\mathbb{I}\partial _{x}-\left( U_{1}\right) _{x}U_{1}^{-1}\qquad \text{%
and\qquad }V_{1}=V+i\left[ \left( U_{1}\right) _{x}U_{1}^{-1},\sigma _{3}%
\right] .
\end{equation}%
We may now simply iterate these equations obtaining%
\begin{equation}
\begin{array}{lll}
U_{2}=\left( L_{1}\Psi (\lambda _{3}),L_{1}\Psi (\lambda _{4})\right) ,~~ & 
L_{2}=\mathbb{I}\partial _{x}-\left( U_{2}\right) _{x}U_{2}^{-1},~~ & 
V_{2}=V_{1}+i\left[ \left( U_{2}\right) _{x}U_{2}^{-1},\sigma _{3}\right] ,
\\ 
U_{3}=\left( L_{2}L_{1}\Psi (\lambda _{5}),L_{2}L_{1}\Psi (\lambda
_{6})\right) , & L_{3}=\mathbb{I}\partial _{x}-\left( U_{3}\right)
_{x}U_{3}^{-1}, & V_{3}=V_{2}+i\left[ \left( U_{3}\right)
_{x}U_{3}^{-1},\sigma _{3}\right] , \\ 
~~~~~\ \ \ \ \ \ ~~~~~\ \ \ \ \ \ \ \ \ \ \vdots & ~\ \ \ \ \ \ \ \ \ \ \ \
\ \ \vdots & ~~\ \ \ \ \ \ \ \ \ \ \ \ \ \ \ \ \ \ \ \vdots \\ 
U_{n}=\left( \mathcal{L}_{n}\Psi (\lambda _{2n-1}),\mathcal{L}_{n}\Psi
(\lambda _{2n})\right) , & L_{n}=\mathbb{I}\partial _{x}-\left( U_{n}\right)
_{x}U_{n}^{-1}, & V_{n}=V_{n-1}+i\left[ \left( U_{n}\right)
_{x}U_{n}^{-1},\sigma _{3}\right] .%
\end{array}
\label{iter}
\end{equation}%
What is left is to specify our original solution $\Psi (\lambda )$. Adopting
the notation from \cite{correa2017}, we abbreviate $\Omega _{i}=\Psi
(\lambda _{i})$ so that at level $n$ of the iteration procedure we have a
set of $2n$ spinors that can be viewed as null states for the intertwining
operator $\mathcal{L}_{n}$ 
\begin{equation}
S_{2n}=\left\{ \Omega _{1},\Omega _{2},\ldots ,\Omega _{2n-1},\Omega
_{2n}\right\} ,\qquad \Omega _{i}=\left( 
\begin{array}{c}
\varphi _{i} \\ 
\phi _{i}%
\end{array}%
\right) ,\quad \lambda _{i}\neq \lambda _{j},  \label{set}
\end{equation}%
i.e. we have $\mathcal{L}_{n}\Omega _{i}=0$ for $i=1,...,2n$.

Having in principle computed $V_{n}$ in an iterative manner, as in indicated
(\ref{iter}), we just need to read off the off-diagonal elements to identify
the new solutions $q_{n}$ and $r_{n}$, because the Darboux-Crum scheme
guarantees that they satisfy the equations (\ref{zero1}) and (\ref{zero2})
when $q$ and $r$ are solutions. These expressions constitute the
multi-soliton solutions we are seeking to construct.

To be explicit, in the first iteration step we have%
\begin{equation}
L_{1}=\mathbb{I}\partial _{x}+\frac{1}{\det W_{1}}\left( 
\begin{array}{cc}
\det D_{1}^{1} & -\det D_{1}^{q} \\ 
\det D_{1}^{r} & \det D_{1}^{2}%
\end{array}%
\right) ,~V_{1}=V_{0}+\frac{2i}{\det W_{1}}\left( 
\begin{array}{cc}
0 & \det D_{1}^{q} \\ 
\det D_{1}^{r} & 0%
\end{array}%
\right) ,
\end{equation}%
where we introduced the matrices%
\begin{equation}
W_{1}=\left( 
\begin{array}{cc}
\varphi _{1} & \phi _{1} \\ 
\varphi _{2} & \phi _{2}%
\end{array}%
\right) ,~D_{1}^{q}=\left( 
\begin{array}{cc}
\varphi _{1}^{\prime } & \varphi _{1} \\ 
\varphi _{2}^{\prime } & \varphi _{2}%
\end{array}%
\right) ,~D_{1}^{r}=\left( 
\begin{array}{cc}
\phi _{1}^{\prime } & \phi _{1} \\ 
\phi _{2}^{\prime } & \phi _{2}%
\end{array}%
\right) ,~D_{1}^{1}=\left( 
\begin{array}{cc}
\varphi _{1}^{\prime } & \phi _{1} \\ 
\varphi _{2}^{\prime } & \phi _{2}%
\end{array}%
\right) ,~D_{1}^{2}=\left( 
\begin{array}{cc}
\varphi _{1} & \phi _{1}^{\prime } \\ 
\varphi _{2} & \phi _{2}^{\prime }%
\end{array}%
\right) .  \label{mat5}
\end{equation}%
From $V_{1}$ we read off the one-soliton solution%
\begin{equation}
q_{1}=q+2\frac{\varphi _{1}^{\prime }\varphi _{2}-\varphi _{1}\varphi
_{2}^{\prime }}{\varphi _{1}\phi _{2}-\varphi _{2}\phi _{1}}\,,\qquad \text{%
and\qquad }r_{1}=r-2\frac{\phi _{1}^{\prime }\phi _{2}-\phi _{1}\phi
_{2}^{\prime }}{\varphi _{1}\phi _{2}-\varphi _{2}\phi _{1}}.  \label{qr1}
\end{equation}%
In a similar fashion we can use now (\ref{iter}) to compute iteratively the
higher order solutions. Remarkably the $n$-solition solutions can be
presented in a closed compact form as 
\begin{equation}
q_{n}=q+2\frac{\det D_{n}^{q}}{\det W_{n}}\,,\qquad \text{and\qquad }%
r_{n}=r-2\frac{\det D_{n}^{r}}{\det W_{n}},  \label{genrqn}
\end{equation}%
with $W_{n}$, $D_{n}^{q}$ and $D_{n}^{r}$ denoting $2n\times 2n$-matrices
generalizing (\ref{mat5}). The determinant of the matrix $W_{n}$ corresponds
to the generalized Wronskian of the set in (\ref{set}) with $n$ columns
containing $\varphi _{i}$, $i=1,...,2n$, and its derivatives and $n$ columns
containing $\phi _{i}$ and its derivatives with respect to $x$ 
\begin{equation}
W_{n}=\left( 
\begin{array}{cccccccc}
\varphi _{1}^{(n-1)} & \varphi _{1}^{(n-2)} & \ldots & \varphi _{1} & \phi
_{1}^{(n-1)} & \ldots & \phi _{1}^{\,\prime } & \phi _{1} \\ 
\varphi _{2}^{(n-1)} & \varphi _{2}^{(n-2)} & \ldots & \varphi _{2} & \phi
_{2}^{(n-1)} & \ldots & \phi _{2}^{\,\prime } & \phi _{2} \\ 
\vdots & \vdots & \ddots & \vdots & \vdots & \ddots & \vdots & \vdots \\ 
\varphi _{2n}^{(n-1)} & \varphi _{2n}^{(n-2)} & \ldots & \varphi _{2n} & 
\phi _{2n}^{(n-1)} & \ldots & \phi _{2n}^{\,\prime } & \phi _{2n}%
\end{array}%
\right) .
\end{equation}%
The matrix $D_{n}^{q}$ is made up of $n-1$ columns containing $\varphi _{i}$
and its derivatives and $n+1$ columns containing $\phi _{i}$ and its
derivatives 
\begin{equation}
D_{n}^{q}=\left( 
\begin{array}{cccccccc}
\phi _{1}^{(n-2)} & \phi _{1}^{(n-3)} & \ldots & \phi _{1} & \varphi
_{1}^{(n)} & \ldots & \varphi _{1}^{\,\prime } & \varphi _{1} \\ 
\phi _{2}^{(n-2)} & \phi _{2}^{(n-3)} & \ldots & \phi _{2} & \varphi
_{2}^{(n)} & \ldots & \varphi _{2}^{\,\prime } & \varphi _{2} \\ 
\vdots & \vdots & \ddots & \vdots & \vdots & \ddots & \vdots & \vdots \\ 
\phi _{2n}^{(n-2)} & \phi _{2n}^{(n-3)} & \ldots & \phi _{2n} & \varphi
_{2n}^{(n)} & \ldots & \varphi _{2n}^{\,\prime } & \varphi _{2n}%
\end{array}%
\right) ,
\end{equation}%
and the matrix $D_{n}^{r}$ is made up of $n+1$ columns containing $\varphi
_{i}$ and its derivatives and $n-1$ columns containing $\phi _{i}$ and its
derivatives%
\begin{equation}
D_{n}^{r}=\left( 
\begin{array}{cccccccc}
\phi _{1}^{(n)} & \phi _{1}^{(n-1)} & \ldots & \phi _{1} & \varphi
_{1}^{(n-2)} & \ldots & \varphi _{1}^{\,\prime } & \varphi _{1} \\ 
\phi _{2}^{(n)} & \phi _{2}^{(n-1)} & \ldots & \phi _{2} & \varphi
_{2}^{(n-2)} & \ldots & \varphi _{2}^{\,\prime } & \varphi _{2} \\ 
\vdots & \vdots & \ddots & \vdots & \vdots & \ddots & \vdots & \vdots \\ 
\phi _{2n}^{(n)} & \phi _{2n}^{(n-1)} & \ldots & \phi _{2n} & \varphi
_{2n}^{(n-2)} & \ldots & \varphi _{2n}^{\,\prime } & \varphi _{2n}%
\end{array}%
\right) .
\end{equation}%
Thus we obtain the $2$-soliton solution from $W_{2}$, $D_{2}^{r}$ and $%
D_{2}^{r}$, the $3$-soliton solution from $W_{3}$, $D_{3}^{r}$ and $%
D_{3}^{r} $, etc. A closed expression for the $\mathcal{L}_{n}$-operator can
be found in \cite{correa2017}$.$

Let us now construct some concrete solutions. First we need to determine $%
\Psi _{1}$ by solving (\ref{ZC}). Specifying the \textquotedblleft seed
functions\textquotedblright\ $q$ and $r$ as $r(x,t)=q(x,t)=0$, taking\ $%
\lambda \rightarrow i\lambda $ the component equations for the two linear
equations $\left( \Psi _{1}\right) _{t}=V\Psi _{1}$ and $\left( \Psi
_{1}\right) _{x}=U\Psi _{1}$ in (\ref{ZC}) decouple into 
\begin{equation}
\left( \varphi _{1}\right) _{x}=\lambda \varphi _{1},~~\left( \phi
_{1}\right) _{x}=-\lambda \phi _{1},~~\left( \varphi _{1}\right)
_{t}=2\lambda ^{2}(i\alpha -2\beta \lambda )\varphi _{1},~~\left( \phi
_{1}\right) _{t}=-2\lambda ^{2}(i\alpha -2\beta \lambda )\phi _{1}.
\label{lin}
\end{equation}%
These equations are easily solved by 
\begin{equation}
\Psi _{1}(x,t;\lambda )=\left( 
\begin{array}{c}
\varphi _{1}(x,t;\lambda ) \\ 
\phi _{1}(x,t;\lambda )%
\end{array}%
\right) =\left( 
\begin{array}{c}
e^{\lambda x+2\lambda ^{2}(i\alpha -2\beta \lambda )t+\gamma _{1}} \\ 
e^{-\lambda x-2\lambda ^{2}(i\alpha -2\beta \lambda )t+\gamma _{2}}%
\end{array}%
\right) ,
\end{equation}%
with constants $\gamma _{1}$,$\gamma _{2}\in \mathbb{C}$. Next we implement
the constraint $r(x,t)=\kappa q^{\ast }(x,t)$, that converts the local
Hirota equation (\ref{1}) into its conjugate (\ref{2}). Given the solution (%
\ref{qr1}) for $r=q=0$ this restriction leads to $\varphi _{2}=\phi
_{1}^{\ast }$, $\phi _{2}=\kappa \varphi _{1}^{\ast }$, so that 
\begin{equation}
\!\Psi _{2}(x,t;\lambda )=\left( 
\begin{array}{c}
\varphi _{2}(x,t;\lambda ) \\ 
\phi _{2}(x,t;\lambda )%
\end{array}%
\right) =\left( 
\begin{array}{c}
\phi _{1}^{\ast }(x,t;\lambda ) \\ 
\kappa \varphi _{1}^{\ast }(x,t;\lambda )%
\end{array}%
\right) =\left( 
\begin{array}{c}
e^{-\lambda ^{\ast }x+2t({\lambda ^{\ast })}^{2}(i\alpha +2\beta \lambda
^{\ast })+\gamma _{2}^{\ast }} \\ 
\kappa e^{\lambda ^{\ast }x-2t({\lambda ^{\ast })}^{2}(i\alpha +2\beta
\lambda ^{\ast })+\gamma _{1}^{\ast }}%
\end{array}%
\right) .
\end{equation}%
Substituting these expressions into (\ref{qr1}) we obtain the one-soliton
solution 
\begin{equation}
q_{1}(x,t)=-\frac{2(\lambda +\lambda ^{\ast })e^{2\lambda x+4\lambda
^{2}t\left( i\alpha -2\beta \lambda \right) +\gamma _{1}-\gamma _{2}}}{%
1-\kappa e^{2(\lambda +\lambda ^{\ast })x+4i\alpha (\lambda ^{2}-({\lambda
^{\ast })}^{2})t-8\beta (\lambda ^{3}+({\lambda ^{\ast })}^{3})t+\gamma
_{1}+\gamma _{1}^{\ast }-\gamma _{2}-\gamma _{2}^{\ast }}}.
\end{equation}%
This solution agrees exactly with the one obtained by means of Hirota's
method in (\ref{oneS}) when we set in there $\lambda \rightarrow 2(\lambda
+\lambda ^{\ast })$, $\mu \rightarrow 2\lambda $, $\gamma \rightarrow \gamma
_{1}-\gamma _{2}$, $\kappa \rightarrow 1$.

In the same way we can construct a $n$-soliton solutions using the set 
\begin{equation}
\tilde{S}_{2n}=\left\{ \Psi _{1}(x,t;\lambda _{1}),\Psi _{2}(x,t;\lambda
_{1}),\Psi _{1}(x,t;\lambda _{2}),\Psi _{2}(x,t;\lambda _{2}),...,\Psi
_{1}(x,t;\lambda _{n}),\Psi _{2}(x,t;\lambda _{n})\right\}
\end{equation}%
with $\lambda _{i}\neq \lambda _{j}$ in the evaluation of the formulae (\ref%
{genrqn}). As we shall discuss below, the solutions to the new non-local
equations are obtained by keeping the same seeds in the constructions of the
wavefunctions $\Psi _{1}$ and by implementing different types of constraints
in the construction of $\Psi _{2}$.

We conclude with a remark on how to obtain degenerate solutions for the
cases with equal eigenvalues as discussed in detail for other models in \cite%
{CorreaFring,cen2016time,CCFsineG}. Instead of considering the set (\ref{set}%
) with $\Omega _{i}$ provided $\lambda _{i}\neq \lambda _{j}$, we need to
implement Jordan states and use the set 
\begin{equation}
\widetilde{S}_{2n}^{\text{deg}}=\left\{ \Omega _{1},\Omega _{2},\partial
_{\lambda }\Omega _{1},\partial _{\lambda }\Omega _{2},\ldots ,\partial
_{\lambda }^{n-1}\Omega _{1},\partial _{\lambda }^{n-1}\Omega _{2}\right\}
,~\Omega _{1}=\left( 
\begin{array}{c}
\varphi _{1} \\ 
\phi _{1}%
\end{array}%
\right) ,\Omega _{2}=\left( 
\begin{array}{c}
\varphi _{2} \\ 
\phi _{2}%
\end{array}%
\right) \,.
\end{equation}%
in the evaluation of the formulae (\ref{genrqn}). Obviously, the
combinations of the two different kind of seeds are also possible giving
rise to new building blocks ${\tilde{S}}_{2n}$ and $\widetilde{S}_{2n}^{%
\text{deg}}$.

\section{The nonlocal complex parity transformed Hirota equation}

In this case the compatibility between the equation (\ref{zero1}) and (\ref%
{zero2}) is achieved by the choice $r(x,t)=\kappa q^{\ast }(-x,t)$. As $x$
is now directly related to $-x$, we expect some nonlocality in space to
emerge in this model.

\subsection{Hirota's direct method}

Let us now consider the new nonlocal integrable equation (\ref{new1}) for $%
\kappa =-1$. We factorize again $q(x,t)=g(x,t)/f(x,t)$, but unlike as in the
local case we no longer assume $f(x,t)$ to be real but allow $%
g(x,t),f(x,t)\in \mathbb{C}$. We then find the identity 
\begin{eqnarray}
&&f^{3}\tilde{f}^{\ast }\left[ iq_{t}\!+\alpha q_{xx}+2\alpha \tilde{q}%
^{\ast }q^{2}-\delta \!\left( q_{xxx}+6q\tilde{q}^{\ast }q_{x}\right) \right]
= \\
&&f\tilde{f}^{\ast }\left[ iD_{t}g\cdot f+\alpha D_{x}^{2}g\cdot f-\delta
D_{x}^{3}g\cdot f\right] +\left( \tilde{f}^{\ast }D_{x}^{2}f\cdot f-2fg%
\tilde{g}^{\ast }\right) \left( \frac{3\delta }{f}D_{x}g\cdot f-\alpha
g\right) .  \notag
\end{eqnarray}%
When comparing with the corresponding identity in the local case (\ref%
{BINLSE}), we notice that this equation is of higher order in the functions
involved, in this case $g$,$\tilde{g}^{\ast }$,$f$,$\tilde{f}^{\ast }$,
having increased from three to four. The left hand side vanishes when the
local Hirota equation (\ref{new1}) holds and the right hand side vanishes
when demanding%
\begin{equation}
iD_{t}g\cdot f+\alpha D_{x}^{2}g\cdot f-\delta D_{x}^{3}g\cdot f=0,
\label{HI1}
\end{equation}%
together with 
\begin{equation}
\tilde{f}^{\ast }D_{x}^{2}f\cdot f=2fg\tilde{g}^{\ast }.  \label{43}
\end{equation}%
We notice that equation (\ref{43}) is still trilinear. However, it may be
bilinearised by introducing the auxiliary function $h(x,t)$ and requiring
the two equations%
\begin{equation}
D_{x}^{2}f\cdot f=hg,\qquad \text{and\qquad }2f\tilde{g}^{\ast }=h\tilde{f}%
^{\ast },  \label{HI2}
\end{equation}%
to be satisfied separately. In this way we have obtained a set of three
bilinear equations (\ref{HI1}) and (\ref{HI2}) instead of two. These
equations may be solved systematically by using in addition to (\ref{expfg})
the formal power series expansion 
\begin{equation}
h(x,t)=\dsum\nolimits_{k}\varepsilon ^{k}h_{k}(x,t).
\end{equation}%
For vanishing deformation parameter $\delta \rightarrow 0$ the equations (%
\ref{HI1}) and (\ref{HI2}) constitute the bilinearisation for the nonlocal
NLSE. As our equation differ from the ones recently proposed for that model
in \cite{stalin2017} we will comment below on some solutions related to that
specific case. The local equations presented in the previous section are
obtained for $\tilde{f}^{\ast }\rightarrow f$, $\tilde{g}\rightarrow g$, $%
h\rightarrow g^{\ast }$ as in this case the two equations in (\ref{HI2})
combine into the one equation (\ref{H2}).

\subsubsection{Two types of one-soliton solutions}

Let us now solve the bilinear equations (\ref{HI1}) and (\ref{HI2}). First
we construct the one-soliton solutions. Unlike as in the local case we have
here several options, obtaining different types. Using the truncated
expansions%
\begin{equation}
f=1+\varepsilon ^{2}f_{2},\qquad g=\varepsilon g_{1},\qquad h=\varepsilon
h_{1},  \label{ans}
\end{equation}%
we derive from the three bilinear forms in (\ref{HI1}) and (\ref{HI2}) the
constraining equations 
\begin{eqnarray}
0 &=&\varepsilon \left[ i\left( g_{1}\right) _{t}+\alpha \left( g_{1}\right)
_{xx}-\delta (g_{1})_{xxx}\right] +\varepsilon ^{3}\left[ 2\left(
f_{2}\right) _{x}\left( g_{1}\right) _{x}-g_{1}\left[ \left( f_{2}\right)
_{xx}+i\left( f_{2}\right) _{t}\right] \right. ~~~~  \label{c1} \\
&&\left. +if_{2}\left[ \left( g_{1}\right) _{t}+i\left( g_{1}\right) _{xx}%
\right] \right] ,~~~~  \notag \\
0 &=&\varepsilon ^{2}\left[ 2(f_{2})_{xx}-g_{1}h_{1}\right] +\varepsilon ^{4}%
\left[ 2f_{2}(f_{2})_{xx}-2(f_{2})_{x}^{2}\right] ,  \label{c2} \\
0 &=&\varepsilon \left[ 2\tilde{g}_{1}^{\ast }-h_{1}\right] +\varepsilon ^{3}%
\left[ 2f_{2}\tilde{g}_{1}^{\ast }-\tilde{f}_{2}^{\ast }h_{1}\right] .
\label{c3}
\end{eqnarray}%
At this point we pursue two different options. At first we follow the
standard Hirota procedure and assume that each coefficient for the powers in 
$\varepsilon $ in (\ref{c1})-(\ref{c3}) vanishes separately. We then easily
solve the resulting six equations by%
\begin{equation}
g_{1}=\lambda \tau _{\mu ,\gamma },\qquad f_{2}=\frac{\left\vert \lambda
\right\vert ^{2}}{(\mu -\mu ^{\ast })^{2}}\tau _{\mu ,\gamma }\tilde{\tau}%
_{\mu ,\gamma }^{\ast },\qquad h_{1}=2\lambda ^{\ast }\tilde{\tau}_{\mu
,\gamma }^{\ast },
\end{equation}%
with constants $\gamma $, $\lambda $, $\mu \in \mathbb{C}$. Setting then $%
\varepsilon =1$ we obtain the exact one-soliton solution 
\begin{equation}
q_{\text{st}}^{(1)}=\frac{\lambda (\mu -\mu ^{\ast })^{2}\tau _{\mu ,\gamma }%
}{(\mu -\mu ^{\ast })^{2}+\left\vert \lambda \right\vert ^{2}\tau _{\mu
,\gamma }\tilde{\tau}_{\mu ,\gamma }^{\ast }}.\quad  \label{sol2}
\end{equation}%
Next we only demand that the coefficient in (\ref{c1})-(\ref{c2}) vanish
separately, but deviate from the standard approach by requiring (\ref{c3})
only to hold for $\varepsilon =1$. This is of course a new option that was
not at our disposal for the standard local Hirota equation, since in that
case the third equation did not exist. In this setting we obtain the
solution 
\begin{equation}
g_{1}=(\mu +\nu )\tau _{\mu ,i\gamma },\qquad f_{2}=\tau _{\mu ,i\gamma }%
\tilde{\tau}_{-\nu ,-i\theta }^{\ast },\qquad h_{1}=2(\mu +\nu )\tilde{\tau}%
_{-\nu ,-i\theta }^{\ast },
\end{equation}%
so that this one-soliton solution becomes%
\begin{equation}
q_{\text{nonst}}^{(1)}=\frac{(\mu +\nu )\tau _{\mu ,i\gamma }}{1+\tau _{\mu
,i\gamma }\tilde{\tau}_{-\nu ,-i\theta }^{\ast }}.  \label{sol1}
\end{equation}

The standard solution (\ref{sol2}) and the nonstandard solution (\ref{sol1})
exhibit qualitatively different behaviour. Whereas $q_{\text{st}}^{(1)}$
depends on one complex spectral and one complex shift parameter, $q_{\text{%
nonst}}^{(1)}$ depends on two real spectral parameters and two real shift
parameters. Hence the solutions can not be converted into each other. Taking
in (\ref{sol2}) for simplicity $\lambda =\mu -\mu ^{\ast }$ the modulus
squared of this solution becomes 
\begin{equation}
\left\vert q_{\text{st}}^{(1)}\right\vert ^{2}=\frac{(\mu -\text{$\mu $}%
^{\ast })^{2}e^{x(\mu +\text{$\mu $}^{\ast })}}{2\cosh \left[ x(\mu -\text{$%
\mu $}^{\ast })\right] -2\cosh \left\{ \gamma +\gamma ^{\ast }+it\left[
\alpha \left[ \mu ^{2}-\left( \mu ^{\ast }\right) ^{2}\right] -\delta \left[
\mu ^{3}-\left( \mu ^{\ast }\right) ^{3}\right] \right] \right\} }.
\label{qst}
\end{equation}%
This solution is therefore nonsingular for $\func{Re}\gamma \neq 0$ and
asymptotically nondivergent for $\func{Re}\mu =0$. We depict a regular
solution in the left panel of figure \ref{regdiv} and observe the expected
nonlocal structure in form of periodically distributed static breathers.

In contrast, the nonstandard solution (\ref{sol1}) is unavoidably singular.
We compute%
\begin{equation}
\left\vert q_{\text{nonst}}^{(1)}\right\vert ^{2}=\frac{(\mu +\nu
)^{2}e^{x(\mu -\nu )}}{2\cosh \left[ x(\mu +\nu )\right] +2\cos \left[
\gamma +\theta +t\left[ \alpha (\mu ^{2}-\nu ^{2})-\delta \left( \mu
^{3}+\nu ^{3}\right) \right] \right] }.  \label{qnonst}
\end{equation}%
which for $x=0$ becomes singular for \textbf{any} choice of the parameters
involved at 
\begin{equation}
t_{\text{s}}=\frac{\gamma +\theta +(2n-1)\pi }{\alpha \left( \nu ^{2}-\mu
^{2}\right) +\delta \left( \mu ^{3}+\nu ^{3}\right) },~~\ \ ~~\ \ \ \ n\in 
\mathbb{Z}\text{.}  \label{rogue}
\end{equation}%
We depict a singular solution in the right panel of figure \ref{regdiv} with
a singularity developing at $t_{\text{s}}\approx -0.689751$. We only zoomed
into one of the singularities, but it is clear from equation (\ref{rogue})
that this structure is periodically repeated so that we can speak of a \emph{%
nonlocal rogue wave} \cite{rogue1,rogue2}.

\FIGURE{ \epsfig{file=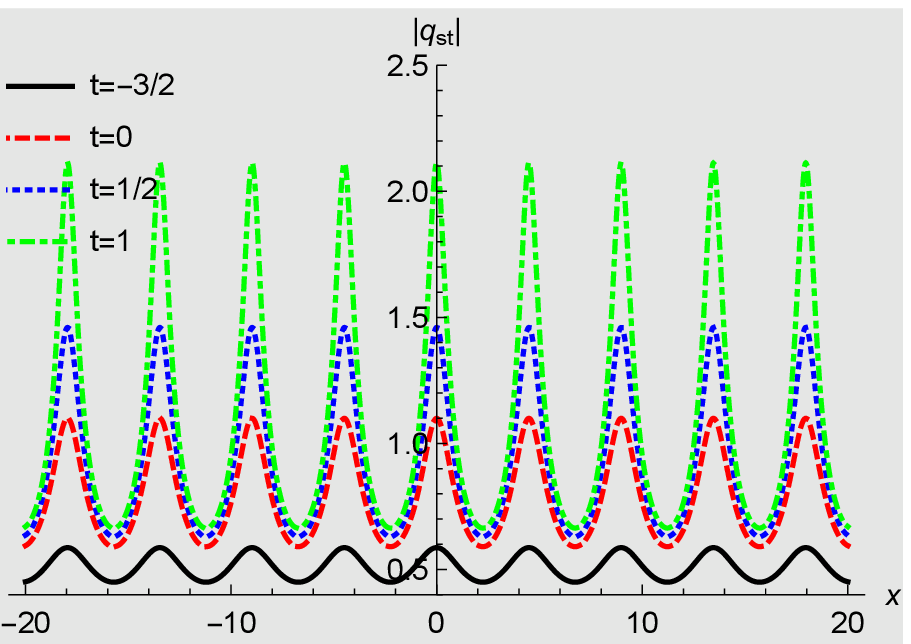,width=7.25cm} \epsfig{file=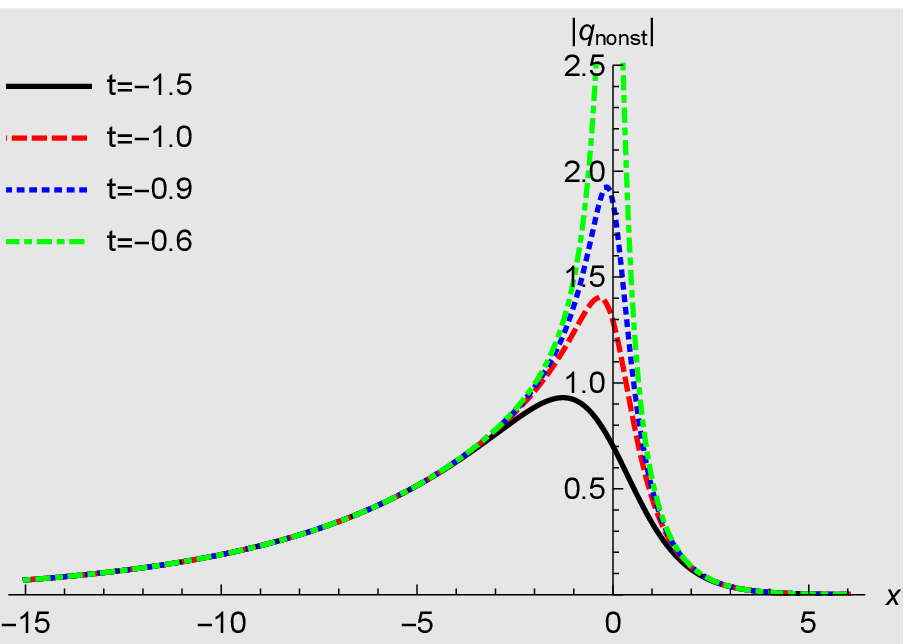,width=7.25cm}
\caption{Nonlocal regular one-soliton solution (\ref{qst}) for the nonlocal Hirota equations obtained from the standard Hirota method at different times for $\alpha = 0.4$, $\delta=0.8$, 
        $\gamma=0.6+i 1.3$ and $\mu=i0.7$, $\lambda = i 1.7$ (left panel). Nonlocal rogue wave one-soliton solution (\ref{qnonst}) for the nonlocal Hirota equations obtained from the nonstandard Hirota method at different times for 
        $\alpha = 0.4$, $\delta=1.8$, $\gamma=0.5$, $\theta=0.1$, $\mu=0.2$ and $\nu=1.2$ (right panel).}
        \label{regdiv}}

Notice that for $\alpha \rightarrow -1$ and $\delta \rightarrow 0$ the
system (\ref{new1}) reduces to the nonlocal NLSE studied in \cite%
{ablowitz2013}. For this case the solution (\ref{sol1}) acquires exactly the
form of equation (22) in \cite{ablowitz2013} when we set $\nu \rightarrow
-2\eta _{1}$, $\mu \rightarrow -2\eta _{2}$, $\gamma \rightarrow \theta _{2}$
and $\theta \rightarrow \theta _{1}$. There is no equivalent solution to the
regular solution (\ref{sol2}) reported in \cite{ablowitz2013}, so that $q_{%
\text{st}}^{(1)}$ for $\delta \rightarrow 0$ is a also new solution for the
nonlocal NLSE.

\subsubsection{The two-parameter two-soliton solution}

As in the local case we expand our auxiliary functions two orders further in
order to construct the two-soliton solution. Using the truncated expansions 
\begin{equation}
f=1+\varepsilon ^{2}f_{2}+\varepsilon ^{4}f_{4},\qquad g=\varepsilon
g_{1}+\varepsilon ^{3}g_{3},\qquad h=\varepsilon h_{1}+\varepsilon ^{3}h_{3},
\end{equation}%
to solve the bilinear equations (\ref{HI1}) and (\ref{HI2}), we find%
\begin{eqnarray}
g_{1} &=&\tau _{\mu ,\gamma }+\tau _{\nu ,\delta },  \label{g1} \\
g_{3} &=&\frac{\left( \mu -\nu \right) ^{2}}{\left( \mu -\mu ^{\ast }\right)
^{2}\left( \nu -\mu ^{\ast }\right) ^{2}}\tau _{\mu ,\gamma }\tau _{\nu
,\delta }\tilde{\tau}_{\mu ,\gamma }^{\ast }+\frac{\left( \mu -\nu \right)
^{2}}{\left( \mu -\nu ^{\ast }\right) ^{2}\left( \nu -\nu ^{\ast }\right)
^{2}}\tau _{\mu ,\gamma }\tau _{\nu ,\delta }\tilde{\tau}_{\nu ,\delta
}^{\ast }, \\
f_{2} &=&\frac{\tau _{\mu ,\gamma }\tilde{\tau}_{\mu ,\gamma }^{\ast }}{%
\left( \mu -\mu ^{\ast }\right) ^{2}}+\frac{\tau _{\nu ,\delta }\tilde{\tau}%
_{\mu ,\gamma }^{\ast }}{\left( \nu -\mu ^{\ast }\right) ^{2}}+\frac{\tau
_{\mu ,\gamma }\tilde{\tau}_{\nu ,\delta }^{\ast }}{\left( \mu -\nu ^{\ast
}\right) ^{2}}+\frac{\tau _{\nu ,\delta }\tilde{\tau}_{\nu ,\delta }^{\ast }%
}{\left( \nu -\nu ^{\ast }\right) ^{2}}, \\
f_{4} &=&\frac{\left( \mu -\nu \right) ^{2}\left( \mu ^{\ast }-\nu ^{\ast
}\right) ^{2}}{\left( \mu -\mu ^{\ast }\right) ^{2}\left( \nu -\mu ^{\ast
}\right) ^{2}\left( \mu -\nu ^{\ast }\right) ^{2}\left( \nu -\nu ^{\ast
}\right) ^{2}}\tau _{\mu ,\gamma }\tilde{\tau}_{\mu ,\gamma }^{\ast }\tau
_{\nu ,\delta }\tilde{\tau}_{\nu ,\delta }^{\ast }, \\
h_{1} &=&2\tilde{\tau}_{\mu ,\gamma }^{\ast }+2\tilde{\tau}_{\nu ,\delta
}^{\ast }, \\
h_{3} &=&\frac{2\left( \mu ^{\ast }-\nu ^{\ast }\right) ^{2}}{\left( \mu
-\mu ^{\ast }\right) ^{2}\left( \nu ^{\ast }-\mu \right) ^{2}}\tilde{\tau}%
_{\mu ,\gamma }^{\ast }\tilde{\tau}_{\nu ,\delta }^{\ast }\tau _{\mu ,\gamma
}+\frac{2\left( \mu ^{\ast }-\nu ^{\ast }\right) ^{2}}{\left( \mu ^{\ast
}-\nu \right) ^{2}\left( \nu -\nu ^{\ast }\right) ^{2}}\tilde{\tau}_{\mu
,\gamma }^{\ast }\tilde{\tau}_{\nu ,\delta }^{\ast }\tau _{\nu ,\delta }.
\label{h3}
\end{eqnarray}%
So that for $\varepsilon =1$ we obtain from (\ref{g1})-(\ref{h3}) the
two-soliton solution%
\begin{equation}
q_{\text{nl}}^{(2)}(x,t)=\frac{g_{1}(x,t)+g_{3}(x,t)}{1+f_{2}(x,t)+f_{4}(x,t)%
}  \label{2nl}
\end{equation}%
As for the one-soliton solution (\ref{sol2}) we recover the solutions to the
local equation by taking $\tilde{\tau}\rightarrow \tau $ and $\mu ^{\ast
}\rightarrow -\mu ^{\ast }$, $\nu ^{\ast }\rightarrow -\nu ^{\ast }$ in the
pre-factors. In figure \ref{nonlocH} we depict the solution (\ref{2nl}) at
different times.

\FIGURE{ \epsfig{file=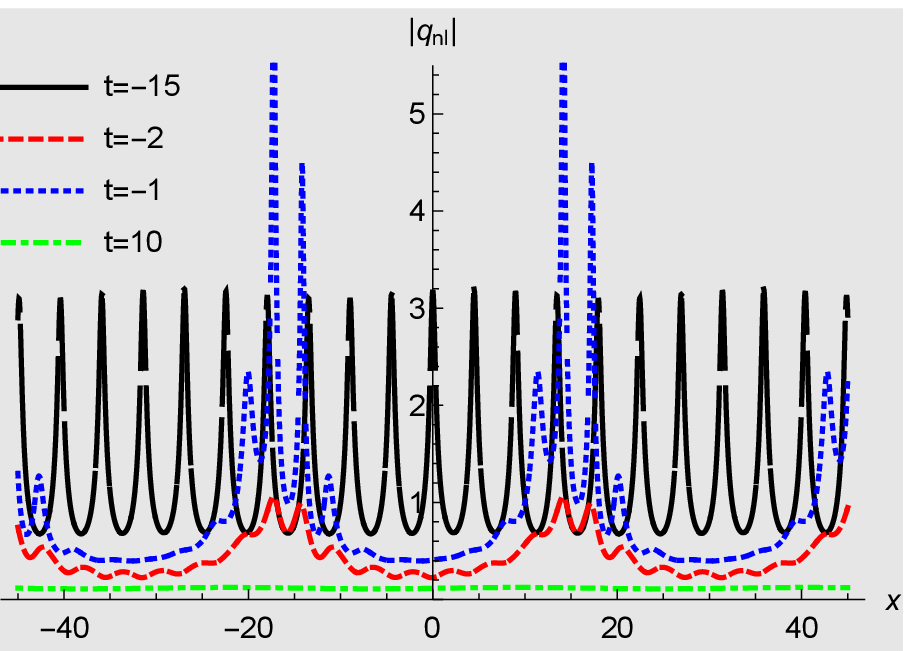,width=7.25cm} \epsfig{file=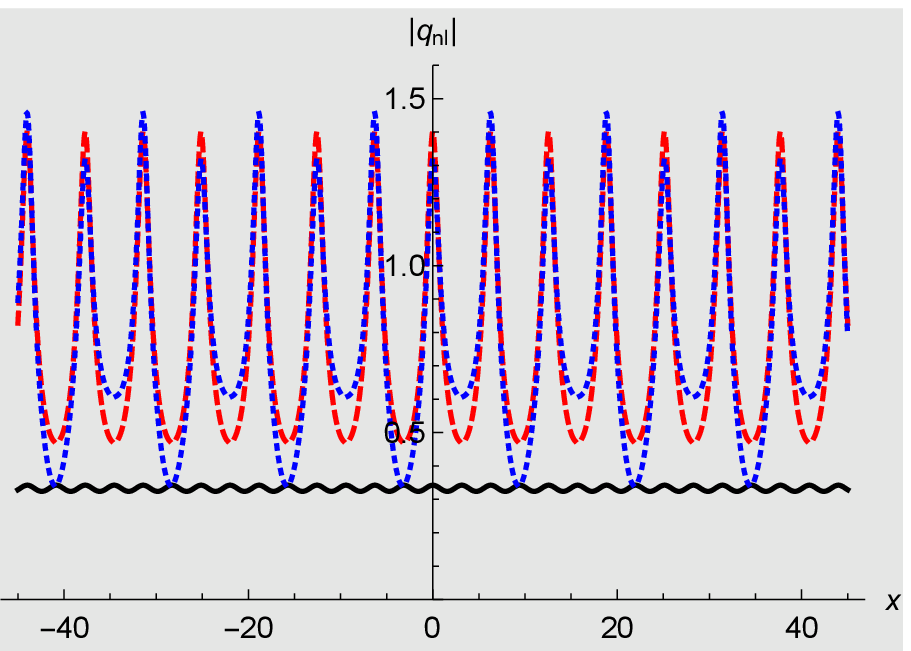,width=7.25cm}
\caption{Nonlocal regular two-soliton solution (\ref{2nl}) for the nonlocal Hirota equations obtained from the standard Hirota method at different times for $\alpha = 0.4$, $\delta=0.8$, 
        $\gamma_1=0.6+i 1.3$, $\mu_1=i0.7$, $\gamma_2=0.9+i 0.7$, $\mu_2=i0.9$ (left panel). Nonlocal regular two one-soliton solution (\ref{sol2}) for the nonlocal Hirota equations $\gamma_1=0.6+i 1.3$, $\mu_1=i0.7$ and $\gamma_2=0.9+i 0.7$, 
$\mu_2=i0.9$ versus the nonlocal regular two-soliton solution (\ref{2nl}) at the same values at time $t=2.5$  (right panel).}
        \label{nonlocH}}

In the left panel we observe the evolution of the two-soliton solution
producing a complicated nonlocal pattern. In the right panel we can see that
at large time the two-soliton solutions appears to be an interference
between two nonlocal one-solitons.

As in the construction for the one-soliton solutions we can also pursue the
option to solve equation (\ref{c3}) only for $\varepsilon =1$ leading to a
second type of two-soliton solutions. We will not report them here, but
instead discuss how they emerge when using Darboux-Crum transformations.

\subsection{Darboux-Crum transformations}

We start again by choosing the vanishing seed functions $q=r=0$ and solve
the linear equations (\ref{ZC}) with $\lambda \rightarrow i\lambda $, i.e. (%
\ref{lin}), with the additional constraint $\beta =i\delta $ by 
\begin{equation}
\tilde{\Psi}_{1}(x,t;\lambda )=\left( 
\begin{array}{c}
\varphi _{1}(x,t;\lambda ) \\ 
\phi _{1}(x,t;\lambda )%
\end{array}%
\right) =\left( 
\begin{array}{c}
e^{\lambda x+2i\lambda ^{2}(\alpha -2\delta \lambda )t+\gamma _{1}} \\ 
e^{-\lambda x-2i\lambda ^{2}(\alpha -2\delta \lambda )t+\gamma _{2}}%
\end{array}%
\right) .  \label{P1}
\end{equation}%
In the construction of $\Psi _{2}$ we implement now the constraint $%
r(x,t)=\pm q^{\ast }(-x,t)$, with $\kappa =\pm 1$, that gives rise to the
nonlocal equations (\ref{new1}) and (\ref{new2}). As suggested from the
previous section we expect to obtain two different types of solutions.
Indeed, unlike as in the local case we have now two options at our disposal
to enforce the constraint. The standard choice consists of taking $\varphi
_{2}=\pm \tilde{\phi}_{1}^{\ast }$, $\phi _{2}=\tilde{\varphi}_{1}^{\ast }$
for complex parameters which is very similar to the approach in local case.
Alternatively we can choose here $\phi _{1}=\tilde{\varphi}_{1}^{\ast }$, $%
\phi _{2}=\pm \tilde{\varphi}_{2}^{\ast }$. Evidently the first equation in
the latter constraint holds when $\gamma _{2}^{\ast }=\gamma _{1}$ in (\ref%
{P1}). It is also clear that the second option is not available in the local
case.

For the first choice we obtain therefore 
\begin{equation}
\tilde{\Psi}_{2}(x,t;\lambda )=\left( 
\begin{array}{c}
\varphi _{2}(x,t;\lambda ) \\ 
\phi _{2}(x,t;\lambda )%
\end{array}%
\right) =\left( 
\begin{array}{c}
\mp \phi _{1}^{\ast }(-x,t;\lambda ) \\ 
\varphi _{1}^{\ast }(-x,t;\lambda )%
\end{array}%
\right) =\left( 
\begin{array}{c}
\mp e^{\lambda ^{\ast }x+2i(\lambda ^{\ast })^{2}(\alpha -2\delta \lambda
)t+\gamma _{2}^{\ast }} \\ 
e^{-\lambda ^{\ast }x-2i(\lambda ^{\ast })^{2}(\alpha -2\delta \lambda
^{\ast })t+\gamma _{1}^{\ast }}%
\end{array}%
\right) ,  \label{P2}
\end{equation}%
with $\lambda $,$\gamma _{1},\gamma _{2}\in \mathbb{C}$ and hence for the
lower sign with (\ref{qr1}) we have 
\begin{equation}
q_{\text{st}}^{(1)}(x,t)=\frac{2(\text{$\lambda $}^{\ast }-\lambda )e^{2%
\text{$\lambda $}^{\ast }x+2i(\text{$\lambda $}^{\ast })^{2}(\alpha -2\delta 
\text{$\lambda $}^{\ast })t-\text{$\gamma $}_{1}^{\ast }+\text{$\gamma $}%
_{2}^{\ast }}}{1+e^{2(\text{$\lambda $}^{\ast }-\lambda )x+4i\left[ \alpha (%
\text{$\lambda $}^{\ast })^{2}-\alpha \lambda ^{2}+2\delta \lambda
^{3}-2\delta (\text{$\lambda $}^{\ast })^{3}\right] t-\gamma _{1}+\gamma
_{2}-\text{$\gamma $}_{1}^{\ast }+\text{$\gamma $}_{2}^{\ast }}}.
\end{equation}%
For the second choice we take $\Psi _{1}(x,t;\mu )$ with $\mu \in \mathbb{R}$
and $\gamma _{2}=\gamma _{1}^{\ast }$ in (\ref{P1}). In this choice the
second wavefunction decouples entirely from the first and we may therefore
also choose different parameters. Again for the lower \ we take 
\begin{equation}
\tilde{\Psi}_{2}(x,t;\nu )=\left( 
\begin{array}{c}
\varphi _{2}(x,t;\nu ) \\ 
\phi _{2}(x,t;\nu )%
\end{array}%
\right) =\left( 
\begin{array}{c}
e^{\nu x+2i\nu ^{2}(\alpha -2\delta \nu )t+\gamma _{3}} \\ 
-e^{-\nu x-2i\nu ^{2}(\alpha -2\delta \nu )t+\gamma _{3}^{\ast }}%
\end{array}%
\right) 
\end{equation}%
and hence (\ref{qr1}) yields 
\begin{equation}
q_{\text{nonst}}^{(1)}(x,t)=\frac{2(\nu -\mu )e^{\gamma _{1}-\gamma
_{1}^{\ast }+2\mu x+4i\mu ^{2}(\alpha -2\delta \mu )t}}{1+e^{2(\mu -\nu
)x+4i(\alpha \mu ^{2}-\alpha \nu ^{2}-2\delta \mu ^{3}+2\delta \nu
^{3})t+\gamma _{1}-\gamma _{1}^{\ast }-\gamma _{3}+\gamma _{3}^{\ast }}}.
\end{equation}%
The $n$-soliton solutions are obtained considering the set 
\begin{equation}
\tilde{S}_{2n}^{\text{st}}=\left\{ \tilde{\Psi}_{1}(x,t;\lambda _{1}),\tilde{%
\Psi}_{2}(x,t;\lambda _{1}),\tilde{\Psi}_{1}(x,t;\lambda _{2}),\tilde{\Psi}%
_{2}(x,t;\lambda _{2}),...,\tilde{\Psi}_{1}(x,t;\lambda _{n}),\tilde{\Psi}%
_{2}(x,t;\lambda _{n})\right\} 
\end{equation}%
or%
\begin{equation}
\tilde{S}_{2n}^{\text{nonst}}=\left\{ \tilde{\Psi}_{1}(x,t;\mu _{1}),\tilde{%
\Psi}_{2}(x,t;\nu _{1}),\tilde{\Psi}_{1}(x,t;\mu _{2}),\tilde{\Psi}%
_{2}(x,t;\nu _{2}),...,\tilde{\Psi}_{1}(x,t;\mu _{n}),\tilde{\Psi}%
_{2}(x,t;\nu _{n})\right\} 
\end{equation}%
with (\ref{P1}) and (\ref{P2}) and the formulae (\ref{genrqn}).

\section{The nonlocal complex time-reversed Hirota equation}

In this case the compatibility between the equations (\ref{zero1}) and (\ref%
{zero2}) is achieved by the choice $r(x,t)=\pm q^{\ast }(x,-t)$ when taking $%
\kappa =\pm 1$. As $t$ is directly related to $-t$, we expect some
nonlocality in time to emerge in this model. Since it is now clear how the
two different types of solutions emerge within the context of the Hirota
method as well as in the application of the Darboux-Crum transformations, we
report here only the latter scenario. Using vanishing seed functions $q=r=0$
we solve the linear equations (\ref{ZC}) with $\lambda \rightarrow i\lambda $%
, $\alpha =i\hat{\delta}$ and $\beta =i\delta $ by 
\begin{equation}
\hat{\Psi}_{1}(x,t;\lambda )=\left( 
\begin{array}{c}
\varphi _{1}(x,t;\lambda ) \\ 
\phi _{1}(x,t;\lambda )%
\end{array}%
\right) =\left( 
\begin{array}{c}
e^{\lambda x-2\lambda ^{2}(\hat{\delta}+2i\delta \lambda )t+\gamma _{1}} \\ 
e^{-\lambda x+2\lambda ^{2}(\hat{\delta}+2i\delta \lambda )t+\gamma _{2}}%
\end{array}%
\right) .  \label{P11}
\end{equation}%
The constraint $r(x,t)=\pm \hat{q}^{\ast }(x,-t)$ in (\ref{qr1}) can be
implemented in two different ways by either taking $\varphi _{2}=\pm \hat{%
\phi}_{1}^{\ast }$, $\phi _{2}=\hat{\varphi}_{1}^{\ast }$ obtaining 
\begin{equation}
\hat{\Psi}_{2}(x,t;\lambda )=\left( 
\begin{array}{c}
\varphi _{2}(x,t;\lambda ) \\ 
\phi _{2}(x,t;\lambda )%
\end{array}%
\right) =\left( 
\begin{array}{c}
\pm \phi _{1}^{\ast }(x,-t;\lambda ) \\ 
\varphi _{1}^{\ast }(x,-t;\lambda )%
\end{array}%
\right) =\left( 
\begin{array}{c}
\pm e^{-\lambda ^{\ast }x-2(\lambda ^{\ast })^{2}(\hat{\delta}-2i\delta
\lambda ^{\ast })t+\gamma _{2}^{\ast }} \\ 
e^{\lambda ^{\ast }x+2(\lambda ^{\ast })^{2}(\hat{\delta}-2i\delta \lambda
^{\ast })t+\gamma _{1}^{\ast }}%
\end{array}%
\right) ,  \label{P22}
\end{equation}%
or $\phi _{1}=\hat{\varphi}_{1}^{\ast }$, $\phi _{2}=\pm \hat{\varphi}%
_{2}^{\ast }$ with $\lambda =\mu \in i\mathbb{R}$, $\gamma _{2}=\gamma
_{1}^{\ast }$ and new constants $\nu \in i\mathbb{R}$, $\gamma _{2}\in 
\mathbb{C}$ so that we have%
\begin{equation}
\hat{\Psi}_{2}(x,t;\nu )=\left( 
\begin{array}{c}
\pm \varphi _{2}(x,t;\nu ) \\ 
\phi _{2}(x,t;\nu )%
\end{array}%
\right) =\left( 
\begin{array}{c}
\pm e^{\nu x-2\nu ^{2}(\hat{\delta}+2i\delta \nu )t+\gamma _{2}} \\ 
e^{-\nu x+2\nu ^{2}(\hat{\delta}+2i\delta \nu )t+\gamma _{2}^{\ast }}%
\end{array}%
\right) .
\end{equation}%
The corresponding one-soliton solutions computed with (\ref{qr1}) are
therefore 
\begin{equation}
q_{\text{st}}^{(1)}(x,t)=\frac{\pm 2(\lambda +\lambda ^{\ast })e^{2\lambda
x+\gamma _{1}+\text{$\gamma $}_{2}^{\ast }}}{e^{2x(\lambda +\lambda ^{\ast
})+4(\lambda ^{\ast })^{2}(\hat{\delta}-2i\delta \lambda ^{\ast })t+\gamma
_{1}+\text{$\gamma $}_{1}^{\ast }}\mp e^{4\lambda ^{2}(\hat{\delta}+2i\delta
\lambda )t+\gamma _{2}+\text{$\gamma $}_{2}^{\ast }}},  \label{kd}
\end{equation}%
and 
\begin{equation}
q_{\text{nonst}}^{(1)}(x,t)=\frac{\pm 2(\mu -\nu )e^{2x(\mu +\nu )+\gamma
_{1}+\gamma _{2}}}{e^{2\mu x+4\nu ^{2}(\hat{\delta}+2i\delta \nu )t+\gamma
_{1}+\text{$\gamma $}_{2}^{\ast }}\mp e^{2\nu x+4\mu ^{2}(\hat{\delta}%
+2i\delta \mu )t+\gamma _{2}+\text{$\gamma $}_{1}^{\ast }}}.
\end{equation}%
\ The nonlocality is now only felt in time for fixed values of $x$, but we
expect to find well localized solutions in space for fixed values of $t$. It
is clear how to compute the $n$-soliton solutions, simply by using the set 
\begin{equation}
\hat{S}_{2n}^{\text{st}}=\left\{ \hat{\Psi}_{1}(x,t;\lambda _{1}),\hat{\Psi}%
_{2}(x,t;\lambda _{1}),\hat{\Psi}_{1}(x,t;\lambda _{2}),\hat{\Psi}%
_{2}(x,t;\lambda _{2}),...,\hat{\Psi}_{1}(x,t;\lambda _{n}),\hat{\Psi}%
_{2}(x,t;\lambda _{n})\right\} 
\end{equation}%
or%
\begin{equation}
\hat{S}_{2n}^{\text{nonst}}=\left\{ \hat{\Psi}_{1}(x,t;\mu _{1}),\hat{\Psi}%
_{2}(x,t;\nu _{1}),\hat{\Psi}_{1}(x,t;\mu _{2}),\hat{\Psi}_{2}(x,t;\nu
_{2}),...,\hat{\Psi}_{1}(x,t;\mu _{n}),\hat{\Psi}_{2}(x,t;\nu _{n})\right\} 
\end{equation}%
with (\ref{P11}) and (\ref{P22}) and the formulae (\ref{genrqn}).

An interesting special case is obtained for $\hat{\delta}=0$, which
correspond to a complex nonlocal time-reverse version of the modified KdV
equation. In this case the solution (\ref{kd}) has no poles for the lower
sign and is asymptotically finite for $t\rightarrow \pm \infty $ as long as $%
\func{Re}\gamma _{1}\neq 0$ and $\func{Re}\gamma _{2}\neq 0$. We depict some
one and two-soliton solutions for the case in figure \ref{nonTwoT}.

\FIGURE{ \epsfig{file=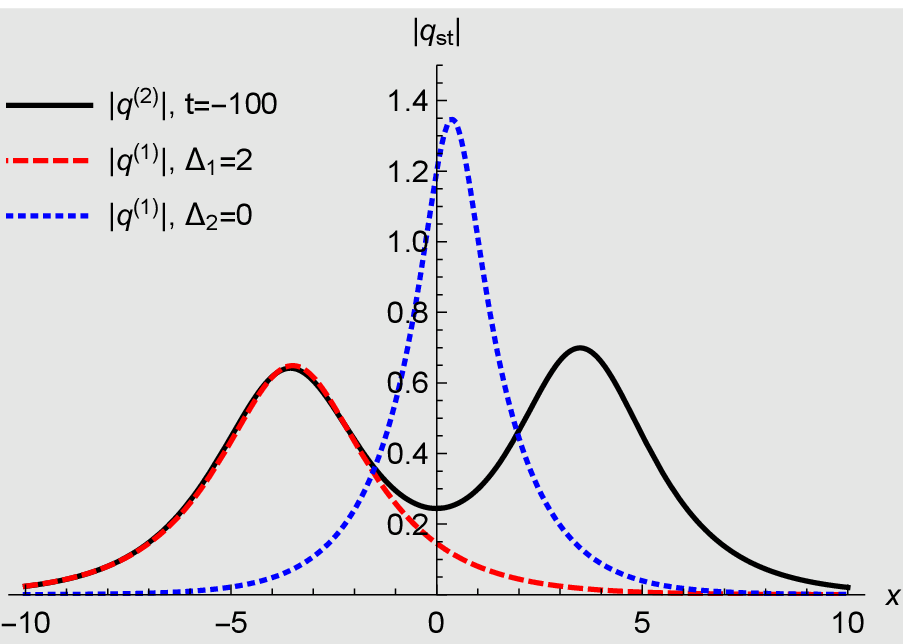,width=4.83cm} \epsfig{file=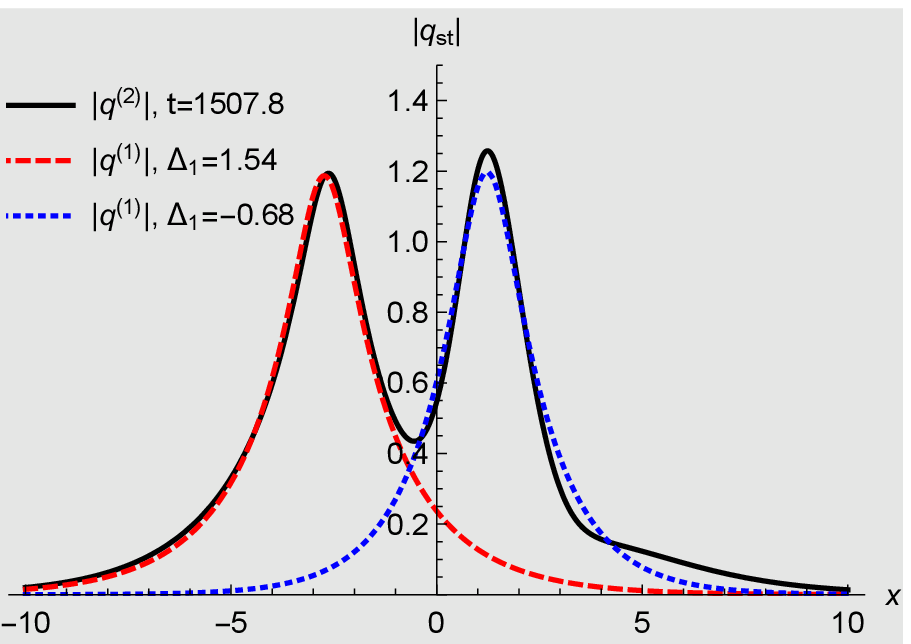,width=4.83cm} \epsfig{file=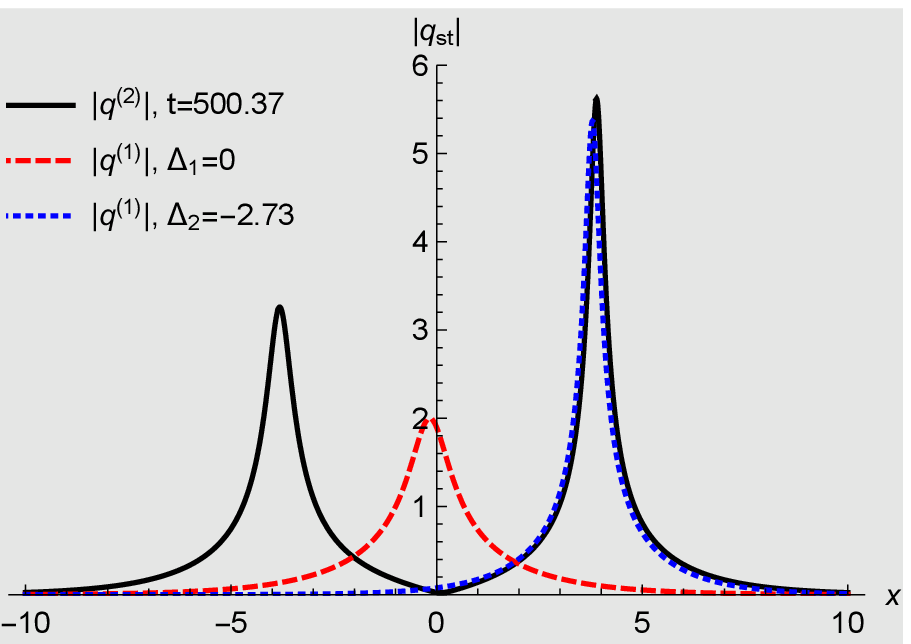,width=4.83cm}
\caption{Modulus of two one-soliton solutions (\ref{kd}) and the corresponding two-soliton solutions for the complex nonlocal time-reverse version of the modified KdV equation at
$\hat{\delta}=0$, $\delta=1.0$ and different values of time. The one-solitons are computed at for $\lambda=0.3$, $\gamma_1=0.3 + \Delta_1$ $\gamma_2=0.2$ and 
 $\lambda=0.4$, $\gamma_1=0.2 + \Delta_2$ $\gamma_2=0.5$. The two-soliton is computed for the same values with $\Delta_1=\Delta_2=0$.}
        \label{nonTwoT}}

We observe from the local nature of the solutions in space and the feature
that the two-soliton solution has one-soliton constituents. As shown in
figure \ref{nonTwoT} when shifting the parameters in the one-soliton
solutions appropriately they match exactly the one-soliton constituents in
the two-soliton solution. However, this only happens at special instances in
time and the solutions will not keep oscillating in synchronicity when time
evolves, even for large values of time.

\FIGURE{ \epsfig{file=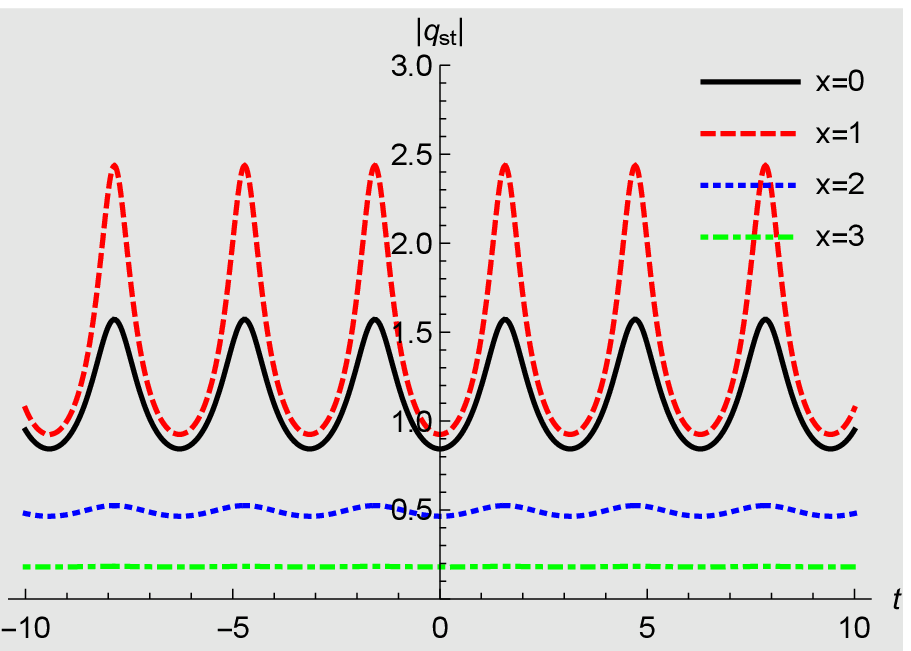,width=7.25cm} \epsfig{file=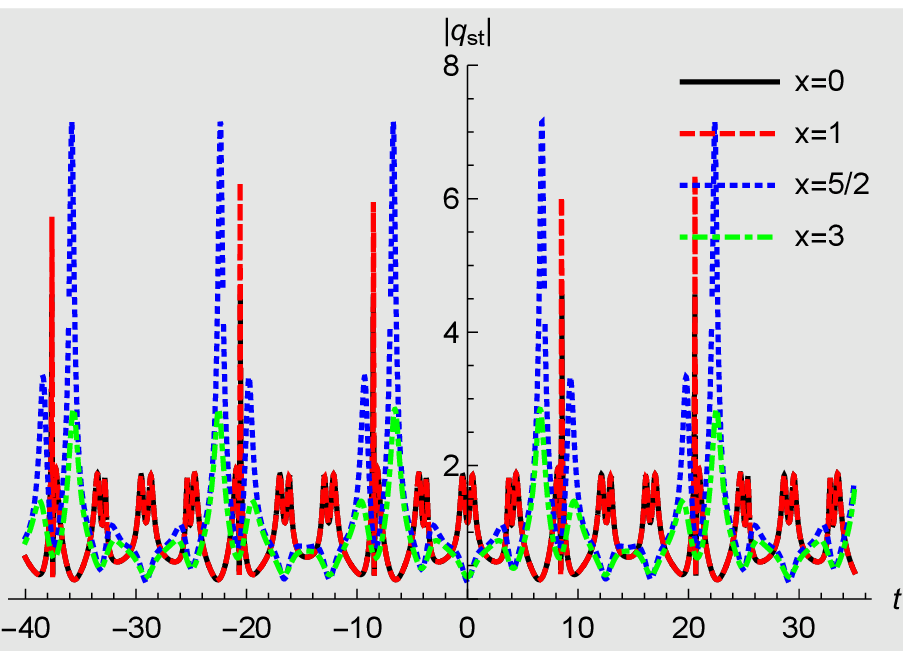,width=7.25cm} 
\caption{Time crystal structures in the complex nonlocal time-reverse version of the modified KdV equation at
$\hat{\delta}=0$, $\delta=1.0$ and different points in space. The one-soliton solutions (\ref{kd}) are computed for $\lambda=0.5$, $\gamma_1=0.2$  and $\gamma_2=0.8$ (left panel).
The two-soliton solutions are computed at for $\lambda_1=0.6$, $\gamma_1=0.2$, $\gamma_2=0.8$, $\lambda_2=0.3$, $\gamma_3=0.2$ and $\gamma_4=0.5$ (right panel).}
        \label{TimeCrystal}}

The nonlocality is now in time, displaying a time crystal \cite%
{Qtimecrystals,Ctimecrystals} like structure as seen in figure \ref%
{TimeCrystal}.

\section{The nonlocal complex $\mathcal{PT}$-transformed Hirota equation}

In this case the compatibility between the equation (\ref{zero1}) and (\ref%
{zero2}) is achieved by the choice $r(x,t)=\pm q^{\ast }(-x,-t)$ when taking 
$\kappa =\pm 1$. As $x$ and $t$ are now directly related to $-x$ and $-t$,
we expect some nonlocality to emerge in space as well as in time. With seed
functions $q=r=0$, $\alpha =i\check{\delta}$, $\check{\delta},\beta \in 
\mathbb{R}$ we solve the linear equations (\ref{ZC}) by 
\begin{equation}
\check{\Psi}_{1}(x,t;\lambda )=\left( 
\begin{array}{c}
\varphi _{1}(x,t;\lambda ) \\ 
\phi _{1}(x,t;\lambda )%
\end{array}%
\right) =\left( 
\begin{array}{c}
e^{\lambda x-2\lambda ^{2}(\check{\delta}+2\beta \lambda )t+\gamma _{1}} \\ 
e^{-\lambda x+2\lambda ^{2}(\check{\delta}+2\beta \lambda )t+\gamma _{2}}%
\end{array}%
\right) .  \label{23}
\end{equation}%
Implementing the constraint $r(x,t)=\pm q^{\ast }(-x,-t)$ in (\ref{qr1}) by $%
\varphi _{2}=\pm \check{\phi}_{1}^{\ast }$, $\phi _{2}=\check{\varphi}%
_{1}^{\ast }$ we obtain 
\begin{equation}
\check{\Psi}_{2}(x,t;\lambda )=\left( 
\begin{array}{c}
\varphi _{2}(x,t;\lambda ) \\ 
\phi _{2}(x,t;\lambda )%
\end{array}%
\right) =\left( 
\begin{array}{c}
\mp \phi _{1}^{\ast }(-x,-t;\lambda ) \\ 
\varphi _{1}^{\ast }(-x,-t;\lambda )%
\end{array}%
\right) =\left( 
\begin{array}{c}
\mp e^{-\lambda ^{\ast }x-2(\lambda ^{\ast })^{2}(\check{\delta}+2\beta
\lambda ^{\ast })t+\gamma _{2}^{\ast }} \\ 
e^{\lambda ^{\ast }x+2(\lambda ^{\ast })^{2}(\check{\delta}+2\beta \lambda
^{\ast })t+\gamma _{1}^{\ast }}%
\end{array}%
\right) ,  \label{34}
\end{equation}%
or $\phi _{1}=\check{\varphi}_{1}^{\ast }$, $\phi _{2}=-\check{\varphi}%
_{2}^{\ast }$ with $\lambda =\mu \in \mathbb{R}$, $\gamma _{2}=\gamma
_{1}^{\ast }$ and new constants $\nu \in \mathbb{R}$, $\gamma _{2}\in 
\mathbb{C}$ we have%
\begin{equation}
\check{\Psi}_{2}(x,t;\nu )=\left( 
\begin{array}{c}
-\varphi _{2}(x,t;\nu ) \\ 
\phi _{2}(x,t;\nu )%
\end{array}%
\right) =\left( 
\begin{array}{c}
-e^{\nu x-2\nu ^{2}(\check{\delta}+2\beta \nu )t+\gamma _{2}} \\ 
e^{-\nu x+2\nu ^{2}(\check{\delta}+2\beta \nu )t+\gamma _{2}^{\ast }}%
\end{array}%
\right) .  \label{12}
\end{equation}%
The corresponding one-soliton solutions computed with (\ref{qr1}) are
therefore 
\begin{equation}
q_{\text{st}}^{(1)}(x,t)=\frac{\pm 2(\lambda -\lambda ^{\ast })e^{\gamma
_{1}+\text{$\gamma $}_{2}^{\ast }+2(\lambda +\lambda ^{\ast })x}}{e^{\gamma
_{1}+\text{$\gamma $}_{1}^{\ast }+4(\lambda ^{\ast })^{2}(2\beta \lambda
^{\ast }+\check{\delta})t+2\mu x}\pm e^{\gamma _{2}+\text{$\gamma $}%
_{2}^{\ast }+4\lambda \mu ^{2}(2\beta \lambda +\check{\delta})t+2\lambda
^{\ast }x}},  \label{qst1}
\end{equation}%
and 
\begin{equation}
q_{\text{nonst}}^{(1)}(x,t)=\frac{2(\nu -\mu )e^{\gamma _{1}-\text{$\gamma $}%
_{1}^{\ast }+2(\mu +\nu )x}}{e^{4\mu ^{2}(2\beta \mu +\check{\delta})t+2\nu
x}+e^{4\nu ^{2}(2\beta \nu +\check{\delta})t+2\mu x}}.  \label{qn1}
\end{equation}%
With (\ref{23}) and (\ref{34}) in the sets 
\begin{equation}
\check{S}_{2n}^{\text{st}}=\left\{ \check{\Psi}_{1}(x,t;\lambda _{1}),\check{%
\Psi}_{2}(x,t;\lambda _{1}),\check{\Psi}_{1}(x,t;\lambda _{2}),\check{\Psi}%
_{2}(x,t;\lambda _{2}),...,\check{\Psi}_{1}(x,t;\lambda _{n}),\check{\Psi}%
_{2}(x,t;\lambda _{n})\right\} 
\end{equation}%
or%
\begin{equation}
\check{S}_{2n}^{\text{nonst}}=\left\{ \check{\Psi}_{1}(x,t;\mu _{1}),\check{%
\Psi}_{2}(x,t;\nu _{1}),\check{\Psi}_{1}(x,t;\mu _{2}),\check{\Psi}%
_{2}(x,t;\nu _{2}),...,\check{\Psi}_{1}(x,t;\mu _{n}),\check{\Psi}%
_{2}(x,t;\nu _{n})\right\} 
\end{equation}%
the $n$-soliton solutions are computed from the formulae (\ref{genrqn}).

As discussed above, the choices $r(x,t)=\pm q(-x,t)$ and $r(x,t)=\pm q(x,-t)$
with real $q$s are less interesting and will therefore not discuss the here.

\section{The nonlocal conjugate $\mathcal{PT}$-transformed Hirota equation}

In this case the compatibility between the equation (\ref{zero1}) and (\ref%
{zero2}) is achieved by the choice $r(x,t)=\kappa q(-x,-t)$ with $\kappa \in 
\mathbb{C}$. As in the previous section we take $q=r=0$, but with no further
restrictions on the parameters involved and solve the linear equations (\ref%
{ZC}) to 
\begin{equation}
\check{\Psi}_{1}(x,t;\lambda )=\left( 
\begin{array}{c}
\varphi _{1}(x,t;\lambda ) \\ 
\phi _{1}(x,t;\lambda )%
\end{array}%
\right) =\left( 
\begin{array}{c}
e^{\lambda x+2\lambda ^{2}(i\alpha -2\beta \lambda )t+\gamma _{1}} \\ 
e^{-\lambda x-2\lambda ^{2}(i\alpha -2\beta \lambda )t+\gamma _{2}}%
\end{array}%
\right) .
\end{equation}%
When implementing the constraint $r(x,t)=\kappa q(-x,-t)$ in (\ref{qr1}) by $%
\varphi _{2}=\kappa \check{\phi}_{1}^{\ast }$, $\phi _{2}=\check{\varphi}%
_{1}^{\ast }$ we obtain a $\Psi _{2}(x,t;\lambda )$ leading to $\det
D_{1}^{q}=\det D_{1}^{r}=0$ so that the standard solution does not exist in
this case. However, implementing $\phi _{1}=i\sqrt{\kappa }\check{\varphi}%
_{1}$, by taking $e^{\gamma _{2}}=i\sqrt{\kappa }e^{\gamma _{1}}$, $\lambda
\rightarrow \mu $ and likewise for $\phi _{2}=-i\sqrt{\kappa }\check{\varphi}%
_{2}$ with new spectral parameters $\lambda \rightarrow \nu $ and shift
parameter $\gamma _{3}=\gamma _{1}$ we obtain 
\begin{equation}
q_{\text{nonst}}^{(1)}(x,t)=\frac{2i(\mu -\nu )e^{2\mu x+4i\mu ^{2}(\alpha
+2i\beta \mu )t}}{\sqrt{\kappa }\left[ 1+e^{2x(\mu -\nu )+\left[ 4i\alpha
(\mu ^{2}-\nu ^{2})+8\beta (\nu ^{3}-\mu ^{3})\right] t}\right] }.
\end{equation}%
We notice that all shift parameters have cancelled and since $x$ and $t$ are
real the solution is in general regular. Interestingly, since the
compatibility requirement between (\ref{zero1}) and (\ref{zero2}) does not
involve a conjugation, we shift formally shift $x$ and $t$ by any complex
value, which means that $q_{\text{nonst}}^{(1)}(x+\Delta _{1},t+\Delta _{2})$
with $\Delta _{1},\Delta _{2}\in \mathbb{C}$ is also a solution that will,
however, in general not satisfy the constraint $r(x,t)=\kappa q(-x,-t)$.
Note that this operation does not constitute a full variable substitution,
i.e. the differentials are not replaced. The $n$-soliton solutions are
obtained by using the set 
\begin{equation}
\check{S}_{2n}^{\text{nonst}}=\left\{ \check{\Psi}_{1}(x,t;\mu _{1}),\check{%
\Psi}_{2}(x,t;\nu _{1}),\check{\Psi}_{1}(x,t;\mu _{2}),\check{\Psi}%
_{2}(x,t;\nu _{2}),...,\check{\Psi}_{1}(x,t;\mu _{n}),\check{\Psi}%
_{2}(x,t;\nu _{n})\right\} ,
\end{equation}%
in the formulae (\ref{genrqn}).

\section{Conclusions}

We exploited various possibilities involving different combinations of
parity, time-reversal and complex conjugation to achieve compatibility
between the two equations (\ref{zero1}) and (\ref{zero2}) resulting from the
zero curvature condition for the Hirota equation. Each possibility
corresponds to a new type of integrable system. Solving these new nonlocal
equations by means of Hirota's direct method we encountered various new
features. Instead of having to solve two bilinear equations, these new
systems correspond to three bilinear equations involving an auxiliary
function. We solved these equations in the standard fashion by using a
formal expansion parameter that in the end can be set to any value when the
expansions are truncated at specific orders. In addition, the new auxiliary
equation allows for a new option for this equation to be solve for a
specific value of the expansion parameter, thus leading to a new type of
solution different from the one obtained in the standard fashion. We also
identified the mechanism leading to this second type of solution within the
approach of using Darboux-Crum transformations. In that context the nonlocal
relations between $q$ and $r$ allow for different options in (\ref{qr1}).

We have found various different type of behaviours. For the local Hirota
equation the sign of the parameter $\kappa $ determines whether the
solutions are regular or singular whereas tuning the spectral parameter can
produce two soliton solutions with a faster one overtaking a slower one, a
head-on collision and, most interestingly, a solution in which one of the
solitons behaves like a defect. The nonlocal complex parity transformed
Hirota equation has two different types of solutions displaying a nonlocal
structure of periodically distributed static breathers or rogue waves. The
nonlocal complex time-reversed Hirota equation possesses regular localized
solutions in space, but is nonlocal in time displaying some time crystal
like structures.

There are various interesting questions left for exploration. Evidently more
concrete scenario for the above cases can be explored and further solutions
may be constructed, for instance by taking different seed function in the
Darboux-Crum transformation etc. We also left aside the study of further
interesting properties, such as degeneracies, time-delays etc., which were
considered in \cite{CenFring,CorreaFring,cen2016time,CCFsineG}. As the approach we followed is general, further new models related to
integrable or even nonintegerable realizations of HNLSE (\ref{HNLSE}) other
than the Hirota equation can be constructed and possibly different types of
systems altogether. The most interesting challenge is to investigate whether
these nonlocal solutions can be realized experimentally.\medskip

\noindent \textbf{Acknowledgments:} JC is supported by a City, University of
London Research Fellowship. FC was partially supported by Fondecyt grant
1171475. AF would like to thank the Instituto de Ciencias F{\'{\i}}sicas y
Matem{\'{a}}tica at the Universidad Austral de Chile for kind hospitality.

\newif\ifabfull\abfulltrue


\end{document}